# Title： Extended Main Sequences in Star Clusters

Chengyuan Li[a,b], Antonino P. Milone[c,d], Weijia Sun[e], and Richard de Grijs[f,g,h]

[a] *School of Physics and Astronomy, Sun Yat-sen University, Daxue Road, Zhuhai, 519082, China*
[b] *CSST Science Center for the Guangdong–Hong Kong–Macau Greater Bay Area, Zhuhai, 519082, China*
[c] *Dipartmento di Fisica e Astronomia "Galileo Galilei", Univ. di Padova, Vicolo dell'Osservatorio 3, Padova, IT-35122, Italy*
[d] *Istituto Nazionale di Astrofisica – Osservatorio Astronomico di Padova, Vicolo dell'Osservatorio 5, Padova, IT-35122, Italy*
[e] *Leibniz-Institut für Astrophysik Potsdam, An der Sternwarte 16, 14482 Potsdam, Germany*
[f] *School of Mathematical and Physical Sciences, Macquarie University, Balaclava Road, Sydney, NSW 2109, Australia*
[g] *Astrophysics and Space Technologies Research Centre, Macquarie University, Balaclava Road, Sydney, NSW 2109, Australia*
[h] *International Space Science Institute–Beijing, 1 Nanertiao, Zhongguancun, Hai Dian District, Beijing, 100190, China*
* Corresponding author: lichengy5@mail.sysu.edu.cn (Chengyuan Li).

## ABSTRACT

Extended main sequences (eMSs) and extended main-sequence turn-offs (eMSTOs) are fascinating phenomena that are routinely observed in star clusters. These phenomena strongly challenge the current canonical "simple stellar population" picture of star clusters, which postulates that star clusters are coeval and chemically homogeneous and can thus be described by a single, unique isochrone. Detections of eMSs and eMSTOs provide valuable insights into stellar physics and the evolution of star clusters. This comprehensive review delves into the observational characteristics, underlying mechanisms, and astrophysical implications of the eMSs and eMSTOs observed in young (less than 600 million years) and intermediate-age (600 to 2000 million years) star clusters. Several scenarios or hypotheses have been proposed to explain these phenomena, including the presence of an age spread, binary interactions, variable stars, and differences in stellar rotation rates. This review discusses the advantages and limitations of current models. Among contemporary models and hypotheses, stellar rotation has been demonstrated as the most plausible mechanism to explain the occurrence of eMSs and eMSTOs. Research on stellar rotation and its connection to eMSs has opened up a myriad of fascinating avenues, such as investigations of the magnetic braking mechanism in stars, searches for tidally locked binary systems in star clusters, and investigations as to whether binary mergers can give rise to massive magnetars. These endeavors have yielded valuable insights and significantly enriched our understanding of stellar astrophysics.

## 1. Introduction

Most stars are formed in molecular clouds [1]. The typical masses of these molecular clouds range from several hundred solar masses to tens of millions of solar masses (giant molecular clouds; GMCs [2]). Given that this mass range is significantly higher than the typical mass of a single star (from 10% of the solar mass to over a hundred solar masses, the solar mass is denoted by $M_\odot$, which is about 1.989×$10^{30}$ kg), it is therefore believed that most, if not all, stars formed in clustered environments [3].

It has long been believed that star clusters are typical representatives of so-called "simple stellar populations" (SSPs [4]). This would imply that all stars in a cluster share the same age and chemical composition. This is so, because the early formation of massive stars in star clusters is accompanied by strong stellar feedback, such as high-energy photon radiation and stellar winds, and they eventually evolve very rapidly into Type II supernovae. Such stellar feedback will rapidly heat and evaporate the gas from the cluster, within a very short time (comparable to the free-fall timescale of molecular clouds [5]), thereby terminating any remaining star-formation activity. This theoretical prediction has been confirmed through a series of observations. For instance, a study of the young massive cluster population in the galaxy M 83 has demonstrated that these clusters expel their initial gas within a relatively short period of time, approximately 4 million years (4 Myr [6]). Similarly, another study conducted of the nearby spiral galaxy NGC 300 has revealed a noticeable spatial decoupling between the molecular gas and high-mass star formation at the scale of GMCs, suggesting a rapid removal of gas in a timeframe of less than 1.5 Myr [7]. Since both M 83 and NGC 300 share similarities with our own Milky Way Galaxy, examining their cluster and star-forming regions can provide valuable insights into the formation and evolution of star clusters in our Galaxy. Because of the so-called "gas expulsion" process just discussed, stars in a cluster can only form within this very short time window before the gas is completely expelled from the cluster (see [8,9]). This results in stars in a cluster often having very small age differences and nearly identical chemical compositions (since, after all, they originate from the same molecular cloud).

However, the consensus that star clusters are SSPs is challenged by studies of star clusters with a full age range. Most clusters older than 2 billion years (2 Gyr) exhibit a significant chemical diversity. Their stars have different compositions of some light elements, including He, C, N, O, Na, and Al, and, in some cases of heavier elements, like iron and s-process elements [10-13]. At the young end, a large number of clusters younger than 2 Gyr harbor an extended main-sequence turn-off (eMSTO) or extended (sometimes split) main sequences (eMSs [14-16]). Although the MSTO regions and upper MSs of these clusters show significant broadening, their lower MSs remain narrow, indicating that the effect of differential extinction on the broadened portion of their MSs is minimal (This indicates that the eMSs only show broadening in the upper section of the MS, and we will later define the upper MS). These eMSTOs and eMSs are in strong conflict with the SSPs scenario, where a unique isochrone can describe the cluster's color–magnitude diagram (CMD). This is the so-called "eMS and eMSTO problem", or "eMS(TO) problem". Although this review only introduces the eMS(TO) problem, we will also discuss the possible connection between the eMS(TO) and multiple stellar populations (MPs) phenomena. For details about the MP problem, we refer the reader to the latest review of [17]. In this review, if we separate eMSTO and eMS, they each represent their own meaning. When we use eMS(TO), it means we are discussing both eMSTO and eMS at the same time.

The eMSTO phenomenon was initially observed in star clusters located in the Large Magellanic Cloud (LMC [18]). By means of observations obtained with the European Southern Observatory's Very Large Telescope (VLT), the LMC cluster NGC 2173 was examined. The observations revealed that the CMD of NGC 2173 exhibited a MSTO region that was wider than anticipated if the cluster were an SSP. Moreover, the analysis revealed that the observed CMD can be matched by a synthetic stellar population characterized by an age distribution of approximately 300 Myr. Subsequent studies, mostly based on *Hubble Space Telescope* (*HST*) photometry, demonstrated that the eMSTO is a common feature for almost all intermediate-age clusters within both the LMC and the Small Magellanic Cloud (SMC), with ages ranging from 1 to 2 Gyr [14,19,20]. These studies indicate that the observed eMSTOs cannot be adequately explained by a traditional SSP model and that the studied clusters seem to have experienced a prolonged star-formation history with a duration of several hundred million years. Subsequent observations revealed that young LMC and SMC clusters, with ages younger than ~600 Myr, also harbor eMSs, and a significant fraction of them even exhibiting split MSs [21–24]. Clearly, these features cannot be explained by SSPs. Thanks to the high-precision astrometric and photometric data released by the *Gaia* mission (DR2, EDR3/DR3), these features have also been found in many Galactic open clusters (OCs [15]). So far, a few hundred clusters younger than 2 Gyr have been found to exhibit eMSTOs or eMSs, with the youngest age (~14 Myr) represented by the *h+χ* Persei double clusters [25]. These clusters comprise a significant fraction (~70%) of the known young star clusters in the Magellanic Clouds (MCs), and this number has also been increasing for OCs in the Milky Way in recent years. The observed CMDs of the LMC clusters NGC 1783 (~1.4 Gyr) and NGC 1856 (~300 Myr), based on HST photometry, are shown in **Fig.1**. It is revealed by this figure that NGC 1783 exhibits an eMSTO, whereas NGC 1856 shows both a split MS and an eMSTO. In contrast, the simulated CMDs for SSPs of the same age and metallicity as NGC 1783 and NGC 1856, which we plotted in **Fig. 1b** and **1d**, respectively, do not show evidence of eMSTOs or eMSs. Similarly to these two clusters, such contradictions between theoretical models and observations are observed in almost all clusters younger than about 2 Gyr.

In this article, we will review recent advancements in research focused on the remarkable eMS(TO)s. Section 2 mainly introduces the characteristics of star clusters with eMS(TO)s, as well as the connections between these features. In Section 3, we discuss the prevailing hypotheses, models, and scenarios proposed to explain eMS(TO)s. We also suggest potential areas for further investigation. Finally, Section 4 presents a comprehensive summary.

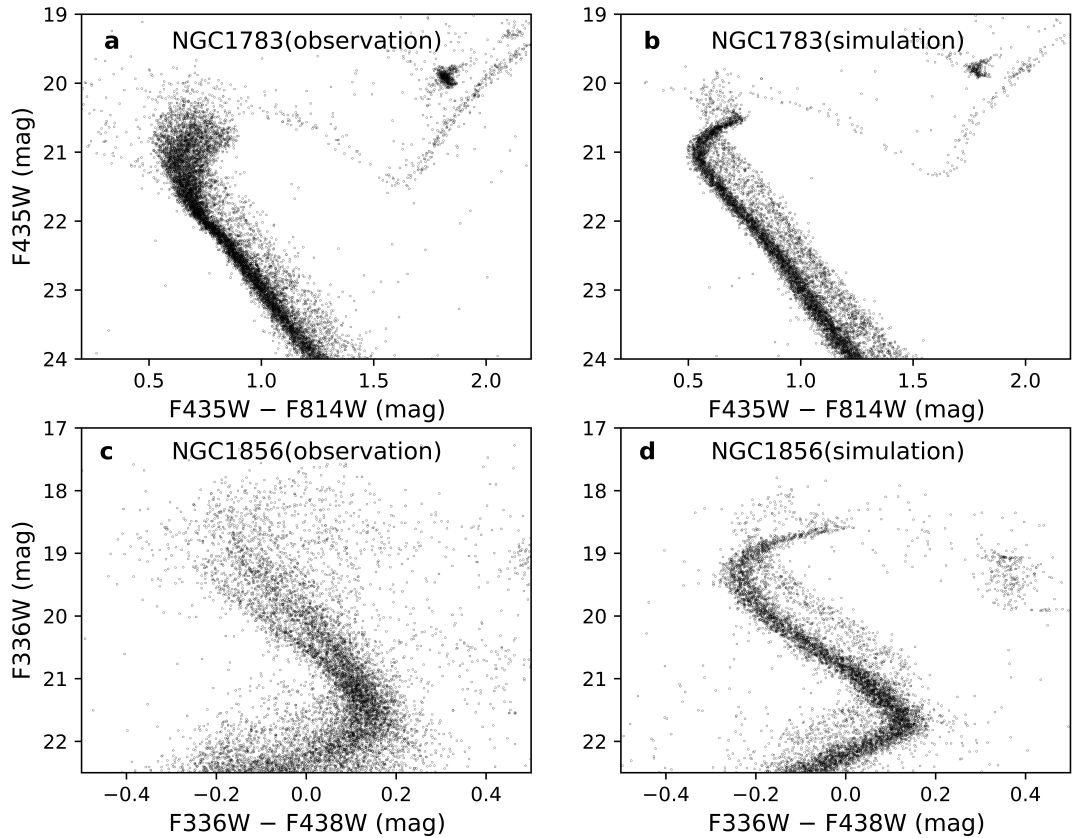

**Fig. 1. Observed and simulated CMDs of clusters NGC 1783 and NGC 1856:** (a) the observed CMD of the ~1.4 Gyr-old cluster NGC 1783. (b) the simulated CMD of a NGC 1783-like SSP cluster. (c) the observed CMD of the ~300 Myr-old cluster NGC 1856. (d) the simulated CMD of a NGC 1856-like SSP cluster.

## 2. Observational characteristics of eMSs

Although eMSTOs and eMSs are commonly observed in, respectively, intermediate-age and young star clusters, these features are not specific to any particular star cluster ages[1]. eMSTOs are not commonly found in young clusters, perhaps simply because these clusters do not have a well-determined MSTO region; instead they may exhibit a broadening of their upper MS region, i.e., eMSs [24]. Sometimes these young clusters even show clear bifurcation among their eMSs (split MSs), like NGC 1856 (see **Fig. 1**). A few intermediate-age clusters—such as NGC 1806, NGC 1846, and NGC 1751—also exhibit two branches in their eMSTO regions [14, 26].

---

[1] In this study, the term "intermediate-age clusters" specifically refers to clusters with ages ranging from 600 to 2000 Myr, whereas clusters with ages younger than 600 Myr are collectively referred to as young clusters. Although the age distributions of MC clusters and OCs in the Milky Way are quite distinct, we do not differentiate between them in this paper.

As examples, we have arranged the CMDs of 20 star clusters in order of their age, from young to old, displayed from left to right and from top to bottom in **Fig.2**. From this figure, it is evident that almost all clusters (except for the oldest cluster, NGC 1978) exhibit an eMS or an eMSTO region. As mentioned, in young clusters, the broadening only shows up in the upper part of the MS above a critical magnitude. As star clusters evolve, this MS broadening gradually shifts towards the MSTO. In the case of the oldest star cluster, NGC 1978, its MSTO magnitude is even fainter than this critical magnitude, resulting in a narrower MS and MSTO region, consistent with the prediction of a SSP. These observations suggest that eMSTOs and eMSs may represent the same underlying physics at different stages of a star cluster's evolution. Clusters displaying eMSTOs may do so because they have reached an age at which the eMS has progressed to the TO stage.

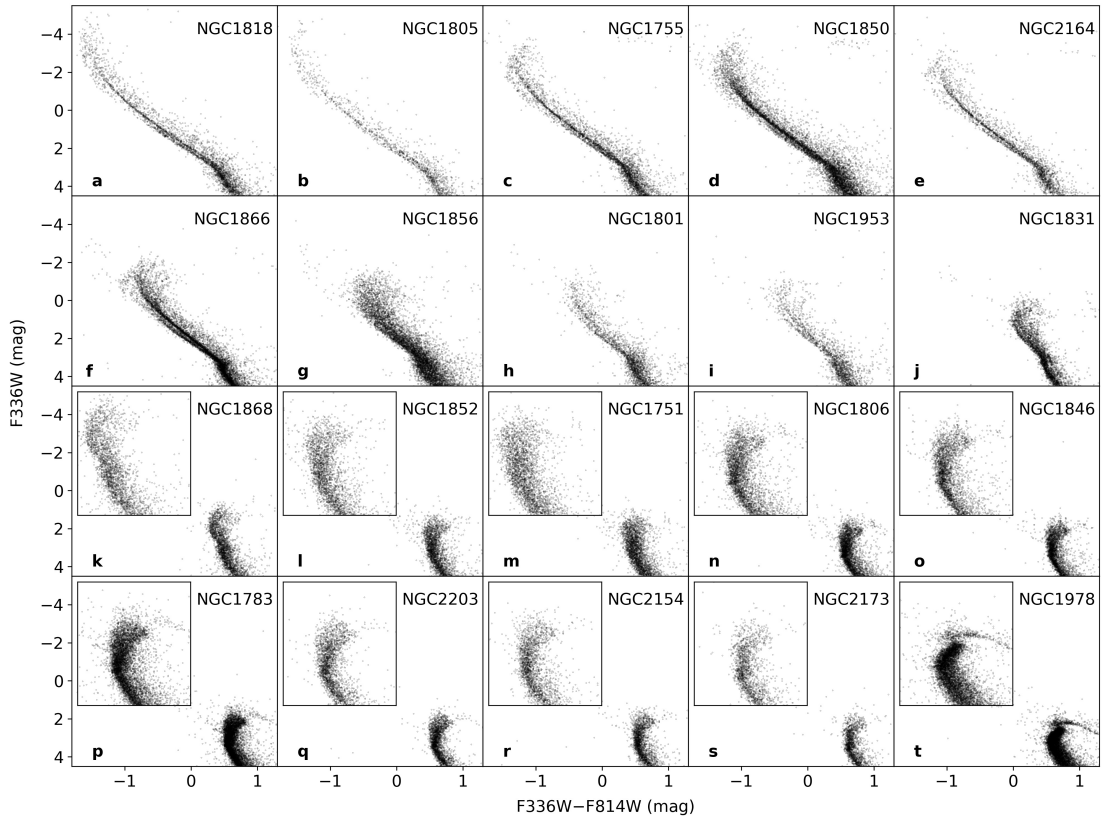

**Fig. 2. CMDs of star clusters of different ages, depicting their MS and MSTO regions, arranged from the top left to the bottom right in order of increasing cluster age (panels a-t, respectively).** The inset zooms in on the MSTO region of select older star clusters (in panels k-t). The data presented in this figure is sourced from [30].

In any star cluster, such a broadening of its characteristic CMD features is only exhibited above a certain position in the MS, sometimes called the MS kink [27]. Therefore, this MS kink defines the upper MS we mentioned earlier, where the broadening only occurs. There is usually no detectable broadening below this critical locus. In other words, the broadening, whether in the MSTO region or along the MS, depends on the corresponding luminosity of the MS, that is, on stellar mass. Several studies suggest that the critical locus may correspond to a stellar mass of approximately $1.5 M_\odot$ [27,28]. As the masses of MSTO stars decrease with increasing cluster age, the broadening will disappear at older cluster ages, as has been confirmed observationally [29,30].

In addition to studying the eMS morphology, research has been carried out to investigate the morphology of other evolutionary sequences within these clusters. The SMC cluster NGC 419 is reported to show the intriguing presence of dual red clumps (RCs [19]), and it also exhibits a significant eMSTOs. A similar pattern is also seen in NGC 411, which exhibits an extended RC region [20]. Contrasting results for another star cluster, NGC 1831, were obtained [31], however. It is found that despite the noticeable eMSTO in this star cluster, its compact RC suggests that it is a cluster composed of a SSP. Several studies have analyzed the subgiant branch (SGB) morphology of MC clusters to investigate whether they meet the expectations of SSPs [32–35]. Intriguingly, despite exhibiting an eMSTO pattern in their CMDs, all clusters examined feature relatively narrow SGBs, which can be better explained by a single-aged stellar population in the presence of the prevailing photometric uncertainties. A distinct evolutionary pattern was observed in the SGB of the SMC cluster NGC 419, with a wide blue section connecting to the broadened eMSTO region, whereas the red section of its SGB gradually narrows towards the bottom of the red-giant branch (RGB [35]). The CMD of NGC 411, showcasing a prime example of a cluster with a remarkable eMSTO, is illustrated in the **Fig. 3a**. However, its SGB is narrow, which can be well explained by a SSP. It has been found that the young star cluster NGC 1818, which exhibits an eMS, features a stellar luminosity function distribution that aligns better with a SSP, particularly at the main-sequence turn-on point (MSTOn [36,37]). In a recent study, it was found that M 37, an OC with an eMS, shows a white dwarf cooling sequence (WDCS) luminosity function that is consistent with the SSP model [38].

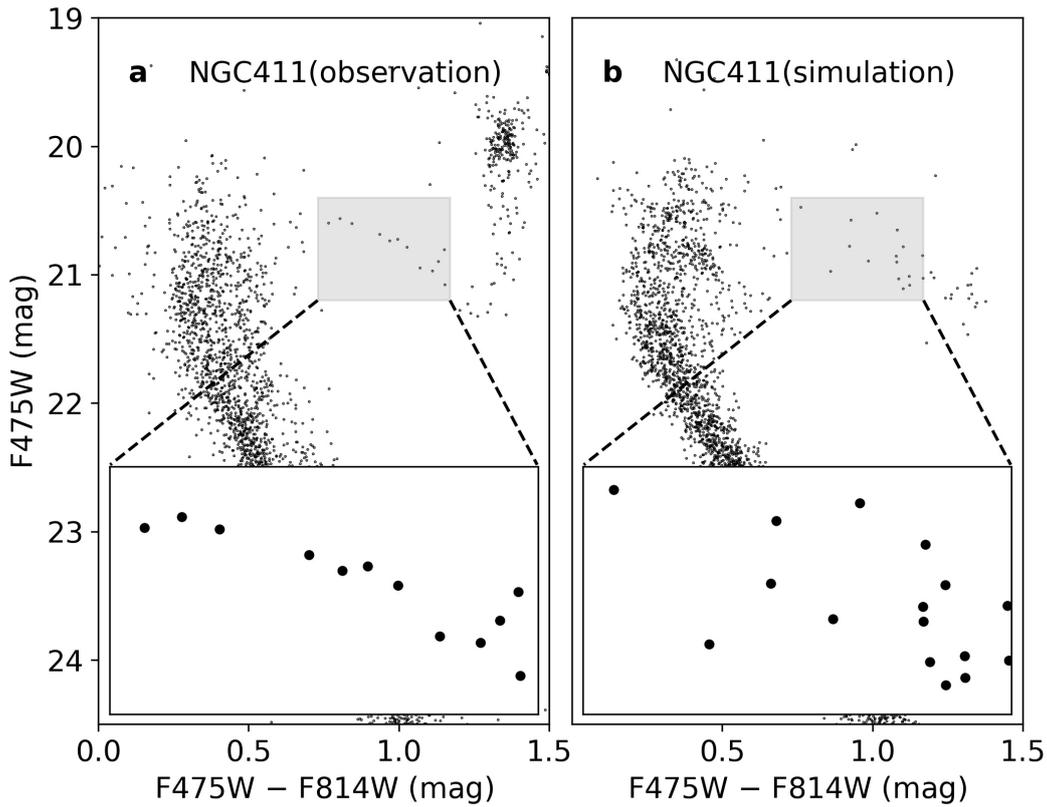

**Fig. 3. Observed and simulated CMDs of the cluster NGC 411.** (a) the observed CMD of cluster NGC 419. (b) a synthetic CMD covering an age range of approximately 700 Myr, proposed to explain the NGC 411 CMD [20]. To emphasize the differences between the two CMDs, we have enlarged the respective SGB regions at the bottom of both panels.

For clusters containing B-type dwarfs that are still located on the MS, a notable fraction of Be stars can be identified through observations employing a filter specifically designed to capture the Hα wavelength (656.28 nm in the rest frame), such as the *HST* F656N filter [29,39]. A Be star is a B-type dwarf with distinctive Balmer emission lines in its spectrum. Since stellar atmospheres are optically thick and cannot produce Balmer emission lines, it is believed that an optically thin circumstellar gas disk surrounding the star is responsible for the observed emission, which is generated by rapid stellar rotation. The fraction of Be stars relative to normal stars depends on spectral type, with the lowest fraction near the A type and the highest fraction towards the MSTO region. In addition to Be stars, recent research has revealed a significant population of "UV-dim" stars in the LMC cluster NGC 1783, which exhibits a clear eMSTO [30]. Because NGC 1783 is too old to have retained any B-type dwarfs, its UV-dim stars are mostly A/F-type stars (and, thus, not Be stars). Their colors are comparable to those of normal A/F-type stars in the F555W–F814W colors defined by the *HST* UVIS/WFC3 filters. However, they appear much redder in the F275W–F438W and F336W–F814W colors, indicating that they are fainter in the UV wavelength range (represented by the F275W and F336W filters) than typical A/F-type stars. As we will elaborate below, in younger clusters a significant number of Be stars also display the analogous UV-dim characteristic, implying a potential relationship between a reduction in UV flux and the presence of a circumstellar gas disk around these stars.

The studies referred to above are all based on *HST* observations of MC clusters. The eMSTO phenomenon was first discovered in the MCs rather than in closer, Galactic OCs. This is because OCs in the Galactic disk are often heavily affected by dust extinction, which obscures their eMS features. Additionally, contamination by field stars in the Galactic disk makes it difficult to determine whether the observed Hertzsprung–Russell diagram features truly originate from cluster member stars. This difficulty was significantly improved only after the second data release of the *Gaia* mission (*Gaia* DR2) in 2018. Also, because the field of view of the *HST* is small, observing closer (and, therefore, larger angular size) OCs requires a significant amount of observation time to combine frames centered on different sky regions, while at the distance of the MCs, the angular size of typical star clusters is small enough to be covered by a single *HST* observation frame.

The majority of young and intermediate-age clusters in the MCs, as well as the ever-increasing number of known OCs in the Milky Way, exhibit eMS(TO)s. However, compared to the clusters in the MCs, the overall broadening of the MSs in Milky Way OCs is not as significant. The MS broadening is predominantly observed in those star clusters within the Milky Way that are rich in stars. Intricate color–magnitude patterns like eMS(TO)s are challenging to discern from the CMDs of sparsely populated OCs. The prevalence of eMS(TO)s among OCs in the Milky Way remains uncertain. Since the discovery of the first Galactic OC exhibiting an eMSTO [40], more than 30 clusters with similar CMD structures have been identified. However, compared to the nearly 4000 known OCs [41], this number evidently falls short of providing a definitive indication as to whether MS broadening dominates across all OCs, not to mention the countless OCs that are believed to have formed, based on theoretical models of the Galactic disk, numbering in the hundreds of thousands [42]. Analyzing such a vast number of star clusters in its entirety would require substantial telescope resources and computational time, leaving us with no choice but to rely on limited statistical samples to estimate the proportion of OCs exhibiting eMS(TO)s. A comprehensive analysis of 72 OCs younger than 2 Gyr in the Milky Way currently suggests that approximately 45% exhibit eMS(TO)s [43]. If this analysis is representative of the overall population of OCs in the Milky Way, it is plausible that a significant proportion of OCs may lack clear evidence of eMS(TO)s (e.g., [44]).

It is worthwhile to investigate the relative spatial distributions of the two MSs composing split MSs, or between different sections of the eMSTOs, which could suggest potential differences in internal kinematics. In several intermediate-age clusters displaying eMSTOs, it has been reported that the brighter section of the MSTO contains stars that are more centrally concentrated than those in the fainter section of the MSTO [45]. In addition, it is also found that these more luminous MSTO stars exhibit a higher degree of central concentration in comparison with RGB and asymptotic giant-branch (AGB) stars. It has thus been proposed that this offers support for the age-spread model, which suggests that the gas released by a first-generation star would initially collect in the cluster center and subsequently trigger the formation of younger and brighter second-generation stars. If bright, blue MS (bMS) stars were to represent a younger stellar population, as suggested for intermediate-age clusters, one would anticipate observing a higher degree of mass segregation among bMS stars in young clusters compared with their presumed older, red-MS (rMS) counterparts. However, no such mass-segregation differences have been detected. In fact, in young clusters with split MSs, most studies have found that the bMS stars are not as centrally concentrated as the majority of MS dwarfs [46,47], or they are at least not significantly different from the rMS stars in terms of their spatial distributions [48]. Another intriguing topic for further exploration pertains to the binary properties of the bMS and rMS. Within these two stellar branches, there may be numerous low-mass-ratio binary stars (hence, their color relative to that of the MS may not have changed significantly). These binary stars, if they are characterized by sufficiently short semi-major axes, may be detectable through measurements of variability in their radial velocities (RVs). Based on a comprehensive analysis using time-domain observations with the VLT/Multi Unit Spectroscopic Explorer (MUSE), a careful investigation of NGC 1850 does not reveal any substantial difference in the binary fraction between the cluster's bMS and rMS [49]. Consequently, that study suggests that the bMS and rMS exhibit comparable binary characteristics within this cluster, at least for the short-period binary systems that can be detected by VLT/MUSE through RV variations.

Undoubtedly, the presence of eMS(TO)s in young and intermediate-age star clusters poses a significant challenge to the standard SSP models. However, it is worth noting that these clusters do not exhibit similarly large variations in stellar chemical abundances as observed in old globular clusters (GCs [50-54]). A Na–O correlation diagram is presented for both GCs and three younger clusters featuring eMS(TO)s in **Fig. 4**. In most GCs, there is a distinct pattern where stars with higher Na abundances tend to have lower O abundances, and *vice versa*. The Na–O anti-correlation is a key characteristic of old GCs and the most significant pattern reflecting the presence of MPs. It is evident that, unlike in GCs, which exhibit a clear Na–O anti-correlation, the Na and O abundances in young star clusters do not show significant differences (within the measurement errors). Although the *HST*'s recent deep photometry of the LMC cluster NGC 1783 implies the possible presence of dispersed nitrogen among its GK-type MS dwarfs [55], additional high signal-to-noise ratio spectroscopic observations are required to verify this assertion.

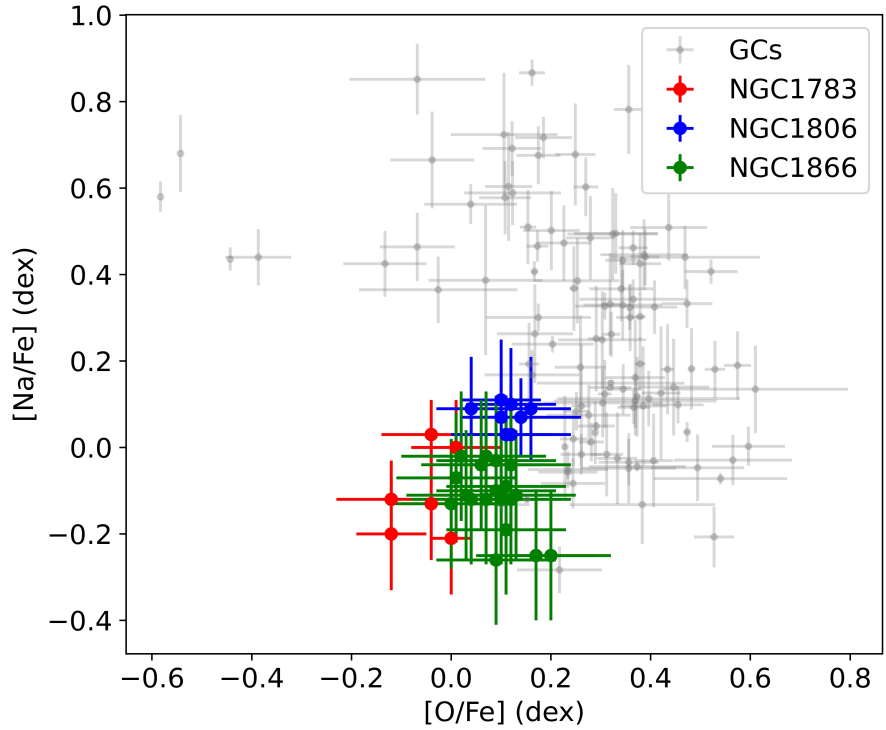

**Fig. 4. Na–O correlation among old GCs in the Milky Way (grey dots) and three clusters of younger and intermediate ages exhibiting eMS(TO)s:** NGC 1783, NGC 1806, and NGC 1866 (red, blue, and green filled circles with error bars, respectively). In contrast to the clear Na–O anti-correlation observed for all Galactic GCs, the three younger clusters with eMSs exhibit a remarkable homogeneity in the Na and O abundances of their member stars. The data used for this figure were obtained from [50–54,56].

## 3. Physical explanations and implications

### 3.1. Proposed explanations

The explanation proposed to account for the origin of eMSTOs was initially quite straightforward. Since the MSTO luminosity is directly determined by the age of an SSP, the presence of eMSTOs seemed to suggest that star clusters could contain stars from multiple generations, that is, characterized by different ages [19, 57–61]. These studies and others, based on direct isochrone fitting or comparisons with synthetic multiply-aged stellar populations, all claimed that an age difference of as much as several hundred million years, with a typical range of 150–250 Myr [14,60] and possibly as large as ~700 Myr [19,20], was required to explain the observed eMSTOs in intermediate-age star clusters. The age spread model also explains the presence of an extended RC in some clusters [19,20], as age spreads will naturally produce a mass range among helium-burning stars.

However, the hypothesis of an age spread directly challenges the conventional understanding of stellar formation models. After all, initial stellar feedback would rapidly disperse the surrounding gas within star-forming clusters (e.g., [62]). The timescale of the gas-expulsion process is comparable to the crossing timescale, or even to the free-fall timescale, which is typically several million years, or shorter. The age-spread model posits that massive clusters have the capacity to undergo multiple epochs of star formation. This scenario is based on the assumption that material processed by stars from an initial population, in the form of stellar ejecta, mixes with a significant amount of primordial (unenriched) gas, thereby giving rise to a second (or multiple) generation(s). For the star-formation process to last for several hundred million years, the dominant contributor to the stellar ejecta would most likely be AGB stars [63]. According to this scenario, a massive cluster (with a total mass of not less than $10,000 M_{\odot}$) could be capable of retaining large amounts of primordial gas or accrete new gas from its surroundings. If this scenario were accurate, the presence of dense pockets in younger clusters (typically between 100 and 500 Myr) would be expected, and these clusters should be actively forming new stars. However, using *Spitzer* 70 μm and 160 μm images and the Australia Telescope Compact Array, combined with Parkes H I survey data, a study focused on identifying H I gas reservoirs in 12 LMC (and one SMC) clusters with ages ranging from 25 to 315 Myr failed to find any compelling evidence of such gas reserves [64]. In addition, a comprehensive study has examined 130 Galactic and extragalactic clusters, spanning a mass range from $10^4$ to $10^8 M_{\odot}$ and ages ranging from 10 to 1000 Myr [65]. This study revealed no signs of ongoing star formation within any of the clusters. The *HST* CMDs of four younger clusters, spanning ages from 20 to 100 Myr, were examined. It was observed that O-type and pre-MS stars were absent in these clusters, suggesting the absence of star formation [24]. Given that the formation environments of intermediate-age clusters with eMSTOs (corresponding to positions at a redshift of $z \sim 0.1$) should be not be very different from those at the present time, the lack of dense gas reservoirs and pre-MS stars in young clusters thus strongly contradicts the age-spread model.

The age-spread model also struggles to explain the structures observed in other parts of the CMD, which often align better with the SSP model. As already mentioned, although the age-spread model explains some clusters with extended RCs, it is challenged by other clusters with a compact RC [31]. Because subgiants represent the next stage after stars leave the MS, the SGB can also be used to constrain the internal age distribution of star clusters. However, star clusters with eMSTOs often have a very narrow SGB, which is inconsistent with the hypothesis of an age spread [32–35]. As an example, despite the presence of an eMSTO and the dual RC in NGC 411, suggesting a potential age dispersion of up to 700 Myr, its narrow SGB clearly contradicts this hypothesis. The synthetic CMD, derived from isochrone fitting to NGC 411 observations and assuming a uniform distribution of different stellar populations with age differences of up to 700 Myr, is displayed in **Fig. 3b.** It is evident that, in contrast to the narrow SGB observed (**Fig. 3a**), a stellar population with such a large age dispersion would exhibit a much broader SGB than what is observed [33].

A potential method to limit the age range of stellar populations in a very young cluster is through the MSTOn, which marks the point in a CMD where the pre-MS merges with the MS. The MSTOn method is an efficient tool for studying the stellar populations of young clusters exhibiting eMSTOs [36]. Employing the MSTOn technique to constrain stellar populations offers the distinct advantage that late-type pre-MS stars are minimally affected by stellar rotation. Conversely, as we will demonstrate below, stellar rotation may significantly alter the morphology of the MSTOs. Using deep imaging from the *HST*, the eMS(TO) phenomenon in the LMC cluster NGC 1818 was investigated. This cluster, with an estimated age of approximately 40 Myr, was analyzed by comparing the observed stellar luminosity function with the best synthetic model. The results showed that the MSTOn morphology of NGC 1818 is consistent with that of a SSP with a relatively small age discrepancy of around 8 Myr [37]. However, focusing solely on the eMSTO feature in the NGC 1818 CMD might lead one to infer that this cluster potentially hosts stars spanning a wide age range, from approximately 30 Myr to nearly 100 Myr.

The WDCS is the process that WDs go through as they cool down over billions of years. As it cools, a WD moves through a predictable and observable sequence of color and magnitude changes, which can be used to determine its age and the time since it evolved into a WD, and it thus can represent the age distribution of a cluster. A recent study investigated the potential age spread in the star cluster M37 by analyzing its WDCS, which exhibits an eMSTO [38]. This study found that the WDCS in M37 lacks a sufficient number of late-type stars, indicating that they have not undergone a sufficiently long cooling period. This is inconsistent with the inferred age dispersion based on the presence of an eMSTO. In the study referenced above, an exhaustive comparison of the WD luminosity function with theoretical models was presented, which led to the exclusion of a significant age difference as the main cause for the eMSTO in the cluster's CMD.

Since the initial discovery of eMS(TO)s in young clusters, a clear correlation has emerged between the putative age spread and the average cluster age. Studies have shown that when the broadening of cluster MSTOs is interpreted as a result of an age spread, this spread is often positively correlated with the age of the cluster. In other words, younger clusters tend to have smaller age spreads, whereas older clusters tend to have larger age spreads [66,67]. Surprisingly, the age spread, if present, appears to reach its maximum in clusters with ages around ~1.5 Gyr and declines swiftly toward older ages [28]. This observation is perplexing, since no reasonable physics can account for why clusters formed approximately ~1.5 Gyr ago would have the longest star-formation history.

There are two proposed scenarios that involve mergers of star clusters with age differences of hundreds of millions of years [68] or of a several hundred million years-old star cluster and a GMC [69]. These scenarios aim to explain the observed dual MSTOs in certain intermediate-age clusters like NGC 1846. The two branches of this cluster's MSTO stars were analyzed, revealing that they share similar metallicities [68]. This contradicts the expectation that a continuous star-formation process would result in a metallicity difference between the two branches. To address this challenge, it is proposed that clusters form in star cluster groups within GMCs and that the cluster-formation process within a cloud can last several hundred million years. This suggested that if the two progenitor stellar components of NGC 1846 were formed in a star cluster group with an approximate age difference of 300 Myr and subsequently merged, this could explain the observed similarity in metallicity between the stars belonging to both branches. The star cluster–GMC merger scenario suggests that clusters with dual MSTOs may have experienced a merger with a star-forming GMC [69]. In this model, secondary star formation is triggered several hundred million years after a cluster's initial formation era. Again, this scenario proposes that the merger event occurred between a cluster and a neighboring GMC, both of which would belong to a "super-cloud" system (or a cloud complex). Consequently, the metallicity difference between the two branches is considered negligible.

All of these models are essentially age-spread models, but they assume that the younger stars observed in star clusters have an external origin, such as accretion through mergers. While these models successfully address the challenge posed by a single star cluster, unable to sustain extended star formation, they face several major problems. First, they cannot explain the prevalence of star clusters with eMSTOs, as mergers between star clusters with significant age differences or between old star clusters and GMCs should not occur so frequently: more than 70% of intermediate-age clusters in the LMC display eMSTOs [14], while binary clusters merely constitute 10% of the current MC cluster population [70]. Additionally, observations have found little difference in the spatial distributions of stars associated with different parts of eMS(TO)s, thus contradicting the expectation that a merger event would result in a more concentrated distribution of younger stars in the core region of the resulting star cluster. Moreover, no evidence of a chemical spread in light elements (such as C, N, or O) has been observed, as predicted by the star cluster–GMC merger scenario. Lastly, these models fail to explain the apparent relationship between the age dispersion in star clusters and the star cluster's average age.

Another hypothesis suggests that the expansion of the MSTO in star clusters may be the result of binary interactions, specifically through merged and interactive binaries [71]. In this model, interactive binaries can produce a large fraction of younger stars (i.e., blue straggler stars), which naturally populate an eMSTO and an extended or dual RC. Although this model does not impose the assumption that star clusters cannot have a long formation history owing to gas expulsion, it cannot adequately explain the observed double MSTO distribution in star clusters like NGC 1846 or NGC 1751. Interactive binaries form a continuous distribution of stars across the MSTO region, and even if all stars are assumed to be members of binary systems, interactive binaries alone cannot fully reproduce the eMSTOs. That is, although interactive binaries may contribute to some overlapping confusion in the MSTO region, they are insufficiently efficient to fully explain the observed eMSTOs [72,73]. Similarly, it has been argued that the binary interaction model cannot explain extended RCs unless it incorporates an unusually high binary frequency [20]. Another limitation is that it fails to account for the absence of eMSTOs in clusters older than ~2 Gyr. This discrepancy remains puzzling, considering that these clusters should also harbor interacting binary systems similar to those in their younger counterparts.

Another scenario attributes the observed eMSTO regions to variable stars [74]. This model is mainly based on the fact that the MSTO region of intermediate-age clusters coincides with the region occupied by δ Scuti stars. δ Scuti stars are a subclass of young pulsating MS stars that exhibit pulsations in radial and non-radial modes. These stars have periods ranging from 0.02 to 0.3 days and can have amplitudes up to 0.9 magnitudes in the *V* band. Due to their presence in the MSTO region, a cluster containing numerous δ Scuti stars will display a broadened MSTO in a single exposure. The advantage of this model is that it does not affect other parts of the cluster CMD, which is consistent with observations [32,33]. In the meantime, it only affects the MSTO region of clusters with ages between 1 and 3 Gyr, which explains why there is no significant broadening of the MS below the MSTO in these clusters. However, similarly to the interactive binary model, the variable-star model also faces the problem of insufficient numbers, as the number of δ Scuti stars is too small to produce a stellar density distribution in the CMD that is almost uniformly broadened at the MSTO. In fact, subsequent observations from the same research group directly refute their previously proposed model [75,76].

A model has been proposed to explain the existence of the eMSs in NGC 1850 and NGC 884, considering differences in interior mixing levels among stars [77]. The authors constructed some "isochrone clouds" (i.e., a group of isochrones) based on the interior mixing profiles of stars with convective cores, which were calibrated using asteroseismology data from isolated Galactic field stars. The measured eMSs were then fitted with these isochrone clouds to estimate the ages and core masses of the stars in both NGC 1850 and NGC 884, assuming coeval populations and fixing the metallicity to the spectroscopically measured values. The correlations between the interior mixing properties of the cluster members and their rotational and pulsational characteristics were subsequently evaluated. They discovered that asteroseismically calibrated interior mixing profiles lead to increased core masses of eMSs stars. In addition, these models account for a significant section of the observed eMSs in the two clusters considered, suggesting the presence of coeval populations of stars with similar ages to those reported in the literature, albeit with considerable uncertainties. According to this model, stars of the same age and mass may exhibit variations in color and magnitude owing to differences in their interior mixing. Specifically, stars characterized by enhanced mixing are expected to appear bluer compared with their less mixed counterparts. This model successfully addresses the need for a prolonged star-formation history in clusters to produce eMS(TO)s by assuming that stars of different colors have different convective core sizes, a phenomenon already observed in numerous early-type Galactic field stars. A potential drawback, however, is that the cause of the interior mixing remains an open question: it could be driven by rotation, pulsations, waves, tides, or binary mergers.

Over a decade ago, the rapid stellar rotation model was first proposed as an explanation for the observed eMSTO [78]. Since then, this model has become the prevailing explanation. According to this model, the eMSTOs seen in intermediate-age clusters can be reproduced by considering the stellar rotation of a SSP. The broadening of the MSTO is driven by "gravity darkening". Gravity darkening causes the equator of a rapidly rotating star to appear darker than its poles. This is due to the variation in surface gravity across the star's surface, which is caused by the centrifugal motion generated by the star's rotation. The centrifugal motion causes the equatorial regions to bulge outward, resulting in a decrease in surface gravity at the equator relative to the poles. Consequently, the equatorial regions of the star are cooler and less luminous than the polar regions, which renders them darker. This effect has been observed in some B- and F-type stars through interferometric imaging [79]. Consequently, stars that possess distinct rates of rotation or rapidly rotating stars that exhibit varying inclination angles will manifest different effective temperatures, giving rise to differences in their colors and fluxes.

However, challenges to the rapid stellar rotation scenario have been raised, highlighting that the effect of internal mixing caused by stellar rotation is not considered in this model [80]. Stellar interior mixing induced by rotation refers to the phenomenon where rotational motion causes the mixing of material within a star. This mixing can have significant and far-reaching consequences for the evolution of stars, including alterations to the structure and properties of the star, such as its fuel consumption rate, energy output, and eventual fate. For example, it has been proposed that the rotational motion of O-, early-B-, and F-type stars could potentially extend their lifetimes by approximately 25%. For more information about how stellar rotation impacts the internal structure of stars through mixing, readers are referred to the classic review by [81]. It has been argued that rotational mixing and gravity darkening will have opposite effects on a cluster's CMD [80]. Whereas gravity darkening will redden and darken a stellar population, their prolonged MS lifetime owing to rotational mixing will make them appear bluer and brighter around the MSTO region. The latter authors showed that these effects will cancel each other out, resulting in a tight MSTO region rather than an eMSTO.

Clearly, the effects of both gravity darkening and rotational mixing can complicate the MSTO's morphology. Therefore, the validity of the refutation strongly depends on the rotation distribution and mixing efficiency set in one's model. It has been highlighted that in the above model, the stellar rotation distribution was characterized by slow speeds, neglecting the presence of a considerable population of stars with exceptionally rapid rotation rates (approaching critical rotational rates) [82]. Consequently, the distribution of stars in the rotating and non-rotating branches during the MS phase in the model exhibited a similar pattern. A study was conducted to quantify the impact of rotation on the morphology of star cluster MSTOs, considering varying levels of mixing efficiency. The results suggested that, with a moderate level of mixing efficiency, an eMSTO could be generated in older clusters (>800 Myr) owing to rotation, but in that case, the eMSTO would disappear in younger clusters [83]. Conversely, a higher level of mixing efficiency would result in a reverse paradigm, where eMSTOs are manifested in younger clusters, whereas older, intermediate-age clusters do not exhibit this feature. A significant implication of their results is that the MSTO area in a cluster depends on its actual age [66]. Other research also indicates that the assumed "age spread" in clusters implied by the eMSTO is dependent on the age of a cluster if it is driven by stars with varying rotation rates [15,67].

The model of stellar rotation has proven successful in resolving numerous issues that are not easily explained by models that consider an age dispersion. This is because the eMSTO is not a consequence of an extended star-formation history, and as a result, it can account for the absence of gas reserves or pre-MS stars in young clusters [24,65]. Additionally, the model can account for the observation that clusters with eMSTOs have narrow SGBs [32–35], since the rotation rates of giant stars would quickly decelerate, resulting in a sequence that is consistent with a SSP (assuming that rotational mixing does not significantly alter the mass distribution of MSTO stars). Considering that rotational mixing can alter the mass distribution of stars at the MSTO, it is theoretically expected that stellar rotation models would also generate a broadened SGB. This issue seems minor, as indicated by the model proposed, which suggests that the most significant variations in mass occur between non-rotating and moderately rotating stars [35]. For stars exhibiting moderate and high rotation rates, the mass differences during their evolution towards the terminal MS are relatively small. Consequently, once they evolve to subgiants, they will exhibit similar luminosities, resulting in a narrow SGB. Therefore, if a substantial majority of stars in these clusters have rotation rates above a certain threshold (although not excessively high), the stellar rotation model can effectively account for the observed narrow SGBs in some star clusters. It is worth noting that the rotational model only imposes limited changes on the appearance of the MS. This explains why pre-MS stars, which are predicted by age-spread models, have not been observed in younger star clusters. Finally, low-mass dwarfs (<~$1.5M_\odot$) will rapidly disperse their initial angular momenta because of magnetic braking [84,85]. This refers to the torque exerted by the stellar magnetic field on ejected matter during stellar evolution, leading to a continuous transfer of angular momentum away from the star. Consequently, the magnetic braking effect suggests that the broadening observed above the MS kink is only apparent, as expected.

The discovery of eMSs initially occurred in LMC clusters, thus posing difficulties in subsequently obtaining high-quality stellar spectra to measure their rotation rates because of their large distances. In 2018, the first direct evidence was provided by measuring the spectra of A- and B-type stars in NGC 1818 [86]. The study revealed that rMS stars displayed higher average rotation rates compared with bMS stars, along with a greater internal dispersion of $v \sin i$. Additionally, Be stars are characterized by the highest $v \sin i$ values and tend to populate the reddest part of the MS. Thanks to *Gaia* DR2, which unveiled the prevalence of eMSs in OCs within our Milky Way, the opportunity arises to conduct direct comparisons of stellar rotational velocities and color properties through high-quality spectroscopic investigations. A study of the Galactic OC NGC 2422 revealed a feature similar to that observed in NGC 1818 [87]. Comparing the multi-band photometry of these stars to synthetic models, the higher dispersion of $v \sin i$ among rMS stars was attributed to unresolved binaries. A strong correlation between stellar colors and rotation rates was discovered in the OC NGC 2287, with bMS stars consistently exhibiting lower $v \sin i$ values compared with rMS stars. Only a few exceptions among the rMS stars displayed slow projected rotation velocities similar to bMS stars [16]. Through careful analysis of the impact of varying rotational rates and inclinations, it was concluded that NGC 2287 contains populations with a bimodal distribution of rotation rates, accompanied by a uniform inclination distribution. Similarly, the eMSTO phenomenon in the Galactic OC NGC 2818 (~800 Myr old) was studied carefully. The results showed that there was a clear difference in the $v \sin i$ values between stars located in the red and blue sections of the eMSTO. Specifically, stars in the red section had significantly higher $v \sin i$ values than those in the blue section. The distribution of color–$v \sin i$ within the cluster was consistent with the theoretical predictions made by stellar models [88]. Using the VLT/MUSE instrument, a team of researchers conducted a meticulous examination of two intermediate-age clusters, NGC 419 (SMC) and NGC 1846 (LMC). Their insightful studies unveiled intriguing findings [89,90]: (i) the broadening of the MSTO is indeed caused by rotation, with stars in the red part of the MSTO exhibiting higher $v\sin i$ and those in the blue section of the MSTO displaying lower $v \sin i$ values. (ii) MSTO stars in NGC 1846 exhibit a bimodal distribution of rotation rates. (iii) Unlike dwarf stars, giant stars exhibit a much narrower rotation distribution, with the majority of evolved giants exhibiting low $v \sin i$. As an example, the CMD of the OC NGC 2287 is presented in **Fig. 5** [16], with each star assigned a color based on their projected surface rotational velocities, $v \sin i$. Notably, stars belonging to the bMS and rMS exhibit distinct rotational velocities, which can be readily discerned from the figure.

Overall, the notion that stellar rotation must play a significant role in the formation of eMS(TO)s is widely accepted. Many of the latest versions of the most commonly used stellar models have incorporated stellar rotation as an optional parameter to provide a more realistic comparison with observations (e.g., MIST models [91]; PARSEC v2.0 [92]; Geneva/SYCLIST database [93,94]). There is only little debate as to whether stellar rotation alone can fully explain all observed details of the eMS(TO)s (e.g., *h+χ* Persei [25]) or if a certain level of age spread is still required (although not as extreme as suggested in the past [95,96]). It is worth noting that uncertainties remain in these stellar evolutionary models, in particular as regards treatments involving processes like turbulence [97]. Therefore, current stellar models that incorporate rotation may not perfectly match the observed data.

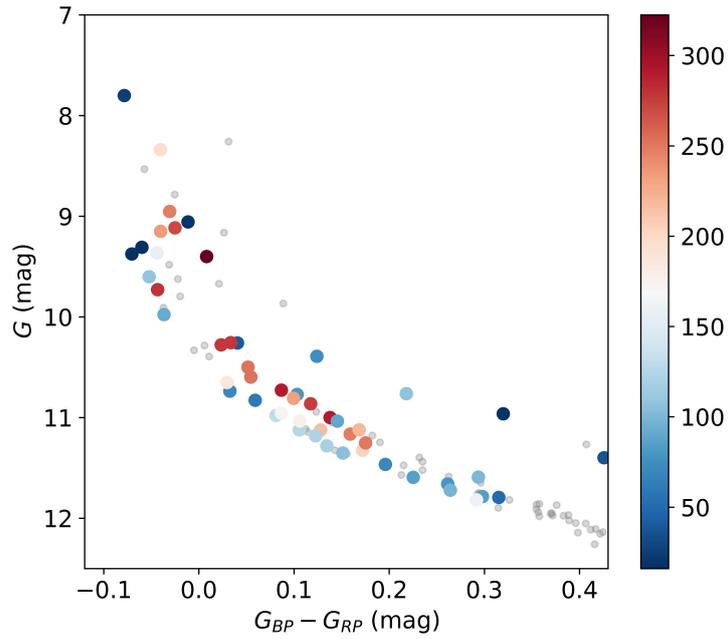

Fig. 5. **CMD of the cluster NGC 2287 observed in the *Gaia* $G_{BP}$, $G_{RP}$, and $G$ bands.** Each star's projected rotational velocity is denoted by a characteristic color (unit: km s$^{-1}$).

### 3.2. Physical implications

The detection of UV-dim stars is shown to further demonstrate the impact of stellar rotation on the morphology of the MSTO. As in NGC 1783, UV-dim stars have been detected in some other clusters (e.g., [98]). It is found that these stars tend to appear in clusters featuring eMS(TO)s and also in younger clusters. The fraction of UV-dim stars is much lower in older clusters. In addition, it has been found that in the young cluster NGC 1850, most B-type UV-dim stars are also Be stars, which thus suggests that dusty decretion disks may account for the UV-flux reduction of these stars. It is important to note that recent research indicates that even B-type stars without emission line features exhibit noticeable attenuation in the ultraviolet band. Therefore, UV-dim stars cannot be considered identical to Be stars [99]. In older clusters—in particular in those older than ~2 Gyr—the majority of MSTO stars manifest themselves as slow rotators owing to the effects of magnetic braking. These clusters thus lack UV-dim stars. Consequently, it has been postulated that UV-dim stars can be classified as "shell stars"—stars with decretion disks—driven by their rapid rotation. Numerical calculations provide further support to this theory, highlighting the significant role of disk self-extinction in generating an eMS(TO) due to variations in stellar rotation. The model suggests that stars with dusty disks, particularly those displaying exceptionally rapid rotation rates, play a crucial role in the formation of the eMS(TO) phenomenon [100]. Additionally, stellar rotation can cause mass loss and form stellar shells, which alters the morphology of the observed stellar luminosity function. Incorporating the impact of disk self-extinction and assuming a canonical mass-luminosity correlation, it was discovered that stars displaying rapid rotation generate an area of increased density along the main sequence. Moreover, if substantial mass loss triggers the magnetic braking effect, it would generate another overdense MS region below the previous one. Consequently, rapid stellar rotation would produce two peaks and a gap in a given magnitude range, corresponding to a stellar mass range of approximately 1.3–1.6$M_\odot$. This phenomenon, referred to as a MS "zig-zag", has been observed in both NGC 1783 and NGC 1978 (**Fig. 6**). However, the use of multi-band photometry has revealed that UV-dim stars have a tendency to occupy the bMS region on the CMD. This region is believed to be dominated by slowly rotating stars, which presents a challenge to the hypothesis mentioned above [99].

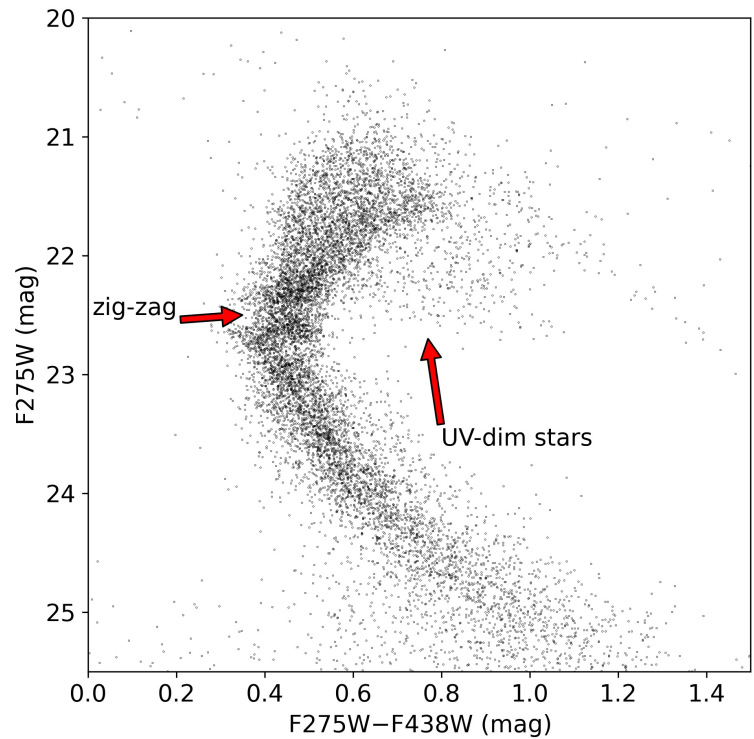

**Fig. 6. The CMD of NGC 1783 seen in the UV (F275W) and *B* (F438W) passbands.** The zig-zag feature and UV-dim stars are marked by horizontal and vertical arrows, respectively.

The origin of the bimodal rotation distribution, which is commonly observed in many clusters with eMSs, remains an open question. It has been proposed that the phenomenon of tidal locking may be responsible for the slow rotation of certain stars, indicating the potential presence of a concealed binary companion [101]. However, if tidal locking in binary systems is the cause of slowly rotating stars, it requires extremely stringent initial parameters for the binary system to produce a bimodal distribution of rotational rates. This is because tidal locking is a gradual, continuous process rather than a bimodal process, so that a continuous distribution of rotation rates is expected in a given star cluster. Tidal locking will only act as a "switch" to suddenly (compared with the age of the star cluster) slow down the rotation of many stars if a considerable fraction of binary systems have tidal-locking timescales that are much shorter than the age of the cluster. Based on these considerations, it has been suggested that a star cluster with an age of 150 million years (similar to the age of NGC 2287) would feature tidal binary systems with initial semi-major axes only several times greater than the radius of the Sun [16]. If such clusters contain so many close binary systems, time-domain spectroscopic observations should reveal a large fraction of RV variable stars. The concept was explored by using NGC 2422 as a testing ground. After conducting observations with the Canada–France–Hawai'i Telescope, no substantial evidence of a high proportion of RV variables was detected. Despite the limited access to observations, with only 2–3 epochs per star, the significance of the results was demonstrated through Monte Carlo models. In conclusion, the number of observed RV variables does not support the existence of a significant number of tidally interacting binaries [102]. In addition, while it is acknowledged that individual observations may not suffice to ascertain the existence of RV variations in particular stars, the study demonstrated that a notable abundance of tidally interacting binaries would result in a broad RV distribution, characterized by a substantial dispersion of RV values among numerous stars. Nevertheless, the observations obtained in this investigation unveiled an exceptionally narrow RV distribution for the stars under scrutiny, effectively negating the likelihood of a significant occurrence of tidally interacting binaries within the cluster [102]. A recent study investigated the possibility of a star cluster similar to NGC 1856 having a significant number of binaries that are tidally locked and located in the blue region of the MS. The results were fascinating, as they suggested that in a young cluster like NGC 1856, tidally locked binaries are likely to have similar masses. This is because in such star clusters, the stellar mass along the MS is already too small to create a strong enough tidal force to quickly lock the primary stars in the binaries. As a result, the secondary star's mass needs to be as close as possible to the mass of the primary star [103]. This is evident from the analytical, approximate expression for the tidal-locking timescale of binary stars [104]:

$$\frac{1}{\tau_{\text{sync}}} = 5 \times 2^{5/3} \left(\frac{GM}{R}\right)^{1/2} \frac{MR^2}{I} q^2 (1+q)^{5/6} E_2 \left(\frac{R}{a}\right)^{17/2}$$

where $\tau_{\text{sync}}$ represents the tidal-locking timescale (also known as the synchronization timescale) of the binary system, and $M$ and $R$ correspond to the mass and radius of the primary star, respectively. The mass ratio, denoted $q$, is defined as the ratio of the mass of the secondary star to the mass of the primary star (thus, $q$ ranges from 0 to 1). It is evident that the tidal-locking timescale of binary stars is highly sensitive to the mass of the primary star, especially when assuming a canonical mass–radius relation for the primary (MS) star of a spectral type earlier than that of the Sun ($R \propto M^{0.8}$). The tidal-locking timescale of a binary system increases rapidly with the decreasing mass of the primary star. Hence, to achieve tidal locking in a cluster within a given age, $q$ must increase correspondingly to counterbalance the effects of the reduced primary star mass. For clusters like NGC 1856, where the primary (MS) stars have relatively low masses, the mass ratio $q$ for binary systems to become tidally locked must be close to unity. However, it is well-established that unresolved binary stars with equal masses will form a parallel sequence to the MS in the CMD, approximately –0.75 magnitudes brighter than the MS. Therefore, tidally locked binaries, if present, cannot account for the observed bMS in NGC 1856. The simulated CMD of a cluster resembling NGC 1856 is shown in **Fig. 7**, where a noteworthy observation emerges: above the MS kick region (at $m_{F336W}$ ~ 21 mag; for reference, see the actual CMD of NGC 1856 in **Fig. 1c**), it becomes apparent that all tidally locked binaries (red crosses) compose a redder sequence compared with their MS counterparts.

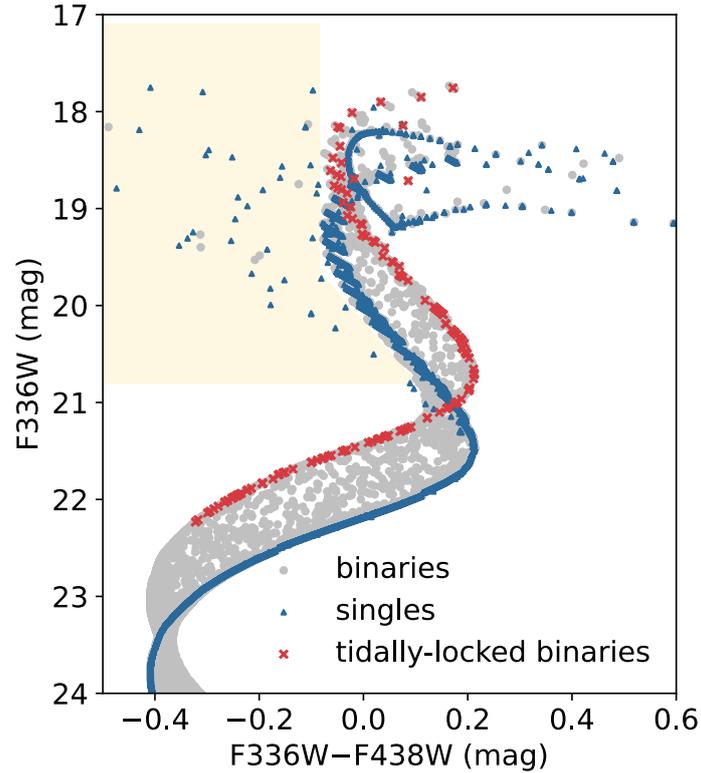

**Fig. 7. Simulated CMD of a NGC 1856-like cluster in the UV (F336W) and *B* (F438W) passbands.** Single stars, binaries, and tidally locked binaries are denoted by blue triangles, grey circles, and red crosses, respectively. The yellow shadow region represents the distribution of blue straggler stars, formed through binary mass transfer or direct mergers.

It has been suggested that the origin of bimodal rotation distributions can be traced back to an early stage, specifically, to a bimodal rotation distribution of pre-MS stars. The model demonstrates that if pre-MS stars with high masses (>1.5 $M_\odot$) were to display a bimodal period distribution and, consequently, a bimodal surface rotation rate, this bimodality could persist throughout their entire MS lifetime [105]. The advantage of this scenario is that many nearby star-forming regions exhibit a bimodal period distribution of their pre-MS stars, although their masses are relatively lower than the mass range we are interested in (0.25–1.0$M_\odot$; e.g., [106]). However, this scenario simply raises another question: What caused the bimodal rotation distribution of pre-MS stars? It has been suggested that interactions between stars and their disks may have a role to play. A star that can retain its protoplanetary disk may be braked, resulting in a slowly rotating star. On the other hand, a rapidly rotating star may not be able to retain its disk for a long time, probably because of interactions with or photo-ionization by a nearby star. This scenario is yet to be examined. Another scenario directly claims that most bMS stars may actually be blue straggler stars [107]. Contrary to our initial intuition, it has been found through recent hydrodynamic calculations that blue stragglers may not necessarily have higher rotational velocities. These calculations indicate that the angular momentum they inherit from the orbital motions of their progenitor binary system can be dissipated rapidly, particularly on a Kelvin–Helmholtz timescale, owing to the emergence of powerful magnetic fields resulting from binary mergers [108]. However, the number of blue straggler stars formed solely through dynamical binary interactions is far from sufficient to account for the observed population of bMS stars ([103]; see Figure 7 and compare it with Figure 1). The model proposes a high rate of mergers within the first few million years following a cluster's formation. A possible explanation for the high formation rate of blue straggler stars is the tidal interactions between binaries and their surrounding circumstellar material. (e.g., [109]). Clearly, an investigation of young embedded clusters focusing on their stellar rotation rates and stellar disks would offer deep insights into both of these scenarios.

The eMS(TO) phenomenon has been used to determine the critical mass below which the magnetic braking effect becomes efficient. As introduced above, only the MS range above this critical mass would exhibit any significant broadening. It is thus expected that only clusters younger than a given age could exhibit eMS(TO)s, because older clusters would have their massive dwarfs (and, thus, rapid rotators) all evolved off the MS. Through an analysis of the significance of eMS(TO)s in clusters of different ages, it has been suggested that the critical age at which magnetic braking becomes ineffective is likely around 2.5 Gyr, which is comparable to the age of the OC NGC 6819. Additionally, the findings indicate that the critical stellar mass for magnetic braking to become ineffective, at solar metallicity, slightly exceeds $1.54 M_\odot$ [28].

It is intriguing to note that star clusters known to host MPs do not exhibit eMS(TO)s. As such, a coexistence of eMSs and MPs within the same star cluster has not been observed [110]. Therefore, a hypothesis has been proposed, suggesting that MPs may arise as a result of some unknown stellar evolutionary effect owing to rotation [111]. If this hypothesis holds true, one would expect to find evidence of MPs among low-mass dwarfs in clusters with eMS(TO)s, although such clusters may not exhibit evidence of MPs among their evolved giant stars. However, two recent pilot studies conducted in NGC 1846 and NGC 419, both example clusters with prominent eMSTOs, have reported no evidence of MPs among late-type dwarfs [112,113]. In addition, as introduced above, eMSs are a common feature of Galactic OCs [15]. The same expectation should also hold for OCs. However, there is no evidence of MPs among old OCs [114,115], which thus directly challenges this hypothesis. Potential validation has been demonstrated by recent findings of a broadening along the lower MS of NGC 1783 (Please note that the MS broadening here is distinct from the eMS we mentioned, as it occurs below the critical mass point, often due to different chemical compositions—such as helium, or C, N, O abundance—of the MS stars constituting this part). The broadening cannot be satisfactorily explained by a SSP, and, instead, an internal dispersion in elemental abundance has been proposed, with a Δ[N/Fe] value of approximately 0.3 dex [55]. It is important to note that further investigation based on direct spectroscopic studies is necessary to confirm this conclusion. Benefiting from extensive multi-epoch observations conducted with the *HST* over a span of more than two decades, researchers have recently achieved a breakthrough in distinguishing stars with a high likelihood of belonging to 58 star clusters in the LMC based on their proper motions [30]. This significant advancement substantially mitigates the interference caused by field stars in the LMC. In the future, using this proper-motion-based, decontaminated data set for MC clusters, statistical analysis may provide insights into whether the broadening of the lower MS in star clusters is related to the presence of MPs.

### 3.3 Future directions

A prospective endeavor that could enhance the precision of both the underlying star-formation history and stellar rotation distribution constraints may revolve around the investigation of youthful star-forming clusters (with ages younger than ~15 Myr). As mentioned, while the eMS(TO) in these clusters may be influenced by a combination of age dispersion and differential stellar rotation, a more accurate understanding of their genuine star-formation histories can be derived from the morphology of their MSTOn. Once we have effectively constrained any internal age spread within these clusters, we can gain deeper insights into the impact of stellar rotation by examining their eMSs. This particular project is eminently feasible for Galactic OCs, thanks to the availability of high-precision photometry from the *Gaia* mission.

To authenticate the theory that bMS stars arise from binary mergers [107], a direct approach would entail evaluating the surface magnetic fields of these stars. Techniques such as Zeeman-effect measurements or polarization analysis can be employed for this purpose. It is worth noting that the study referenced above focused on a specific star cluster within the LMC, and these measurement methods require high signal-to-noise ratios, making it challenging to directly validate the clusters of interest. However, investigating OCs in the Milky Way that exhibit a bifurcated MS could provide a promising avenue for conducting the necessary measurements. In particular, a remarkable bifurcation in the MS was detected within the OC NGC 2287 [16], characterized by a well-defined bimodal rotation distribution. Direct measurements of the magnetic field of bMS stars in NGC 2287 would be crucial for a thorough examination of the proposed hypothesis.

The eMS(TO)s characteristics in star clusters offer an opportunity to investigate the mechanism of magnetic braking, particularly in environments defined by different metallicities. The presence of a strong magnetic field on the surface of a star is determined by whether its core is dominated by convection or radiation, which in turn depends on the rate of the central CNO-cycle reactions. This is so, because the CNO cycle has a much greater temperature sensitivity, which thus results in a higher temperature in the core region, in turn leading to a convective core. Since carbon, nitrogen, and oxygen are metals that catalyze the CNO cycle, a star's metallicity influences the CNO reaction rates. In the meantime, the higher the metallicity, the stronger the stellar winds, resulting in a greater loss of angular momentum. Therefore, it is worth investigating whether stellar metallicity also affects the critical mass for magnetic braking. Recent investigations have revealed a crucial threshold, estimated to be around ~$1.54 M_\odot$, below which the magnetic fields of stars start to gain significance [28]. Nevertheless, it is worth noting that this critical mass might vary among clusters located in the MCs, where metallicities are below solar levels. As an illustration, it has been shown that a lower limit of roughly ~1.45 $M_\odot$ could be relevant for NGC 1831, beyond which the impact of stellar rotation becomes noteworthy [27]. Exploring OCs in the Milky Way and young to intermediate-age clusters in the MCs allows us to gain insights into variations of the stellar surface magnetic fields with stellar mass for both solar and subsolar metallicities. Consequently, such an investigation would help us comprehend the interplay between metallicity-driven stellar-wind processes and stellar surface magnetic fields in removing angular momentum from the stellar surface.

Currently, UV-dim stars have only been detected in a few star clusters in the LMC [98]. Is the presence of these stars a common feature? Do these stars exist from the birth of the star clusters, or do they require a certain amount of time to form? Studies focusing on these topics are currently lacking. A large-scale UV–optical survey of star clusters would greatly contribute to our understanding of these questions. In this regard, the next-generation *China Space Station Telescope* (*CSST*) will be equipped with similar observational wavelengths and filter combinations as the *HST* (e.g., [116,117]), but boasting a field of view about ~7–300 times larger. The operation of the *CSST* undoubtedly offers an opportunity for future research into the origins of UV-dim stars. Stellar metallicity may influence the formation of these UV-dim stars, assuming that their characteristics are influenced by stellar rotation. Lower metallicity levels result in reduced mass-loss rates, allowing stars to maintain higher rotation rates and consequently form more shell stars. A comparative analysis of UV-dim stars in the MCs and the Milky Way would provide valuable insights into this matter.

Assuming that stellar rotation can fully account for the appearance of eMSs in star clusters, an intriguing question arises: why is the proportion of OCs in the Milky Way exhibiting eMS(TO)s significantly smaller compared to the young clusters in the MCs? Currently, there is no research available that provides an explanation. Here, we propose some exploratory hypotheses: one possible explanation could be the difference in the instruments used for observing the Milky Way and MC star clusters. For the MCs, it has been observed that their young clusters exhibit a significant broadening of the MS in UV passbands, whereas the eMS(TO) pattern is weak in optical bands and has thus long been overlooked. In addition, we have realized that the majority of OCs that do not exhibit any significant eMS are those containing fewer stars. A possible explanation for this could be that these clusters have too few stars to form broadening structures in their MSs. To illustrate this, in Figure 8 we present three Galactic OCs. The top left cluster, NGC 3532, comprises approximately 2000 stars (observational data from *Gaia* DR3). The evident eMS(TO) in this rich cluster can be easily detected from *Gaia*'s high-precision photometry. Conversely, the top middle and top right clusters, Blanco 1 and IC 2391, respectively, include modest star counts of around 200 to 300 members. A visual inspection reveals that the latter two clusters lack any discernible indications of an eMS(TO)s. The bottom left panel of Figure 8 shows the CMD of the LMC cluster NGC 1866, observed at UV (F336W) and optical (F814W) wavelengths. The presence of an eMS(TO) in NGC 1866 is evident. However, when we observe this cluster in optical passbands (bottom middle panel, using *HST*'s F438W and F555W filters, similar to *Gaia*'s *BP* and *RP* filters), the eMS(TO) becomes less significant. Moreover, if we randomly select a subsample of ~200 stars from the NGC 1866 population, any signature of eMS(TO) becomes entirely undetectable. We therefore speculate that the differences in observed passbands and sample sizes may be one of the reasons for the differences in the eMS(TO)s between Galactic OCs and MC clusters.

If the observed differences between Galactic OCs and MC clusters are physically real, that could indicate that stellar rotation may be correlated with the environment, such as with the metallicity at the time of cluster formation. In fact, not only do Galactic OCs exhibit distinct signatures of their eMS(TO)s compared with MC clusters, but there are also differences in the eMS(TO)s among clusters within the LMC and SMC. It has been discovered that two clusters in the SMC have a lower fraction of bMS stars compared with clusters in the LMC [29]. Once again, we anticipate that the *CSST* will play a crucial role in this topic. This is not only due to its unique UV-band imaging capabilities but also because of its large field of view, enabling the observation of a greater number of star clusters within the local group.

The bimodal distribution of rotational velocities is also observed among field stars in the Milky Way. Through the application of the LAMOST medium-resolution spectroscopic survey, it was discovered that this bimodal rotation distribution is primarily applicable to stars with masses exceeding $~2.5M_\odot$. This observational finding exhibits slight disparities when compared to star clusters. In star clusters, a bifurcation in the MS occurs above a critical mass (for dwarfs earlier than F-type), indicating the presence of a bimodal distribution of rotational velocities even for stars with relatively low masses ($>1.5M_\odot$). Moreover, it has been observed that this bimodal rotational distribution ceases to exist among field stars with relatively low metallicities ([M/H] < –0.2 dex), which contrasts with the observed split MSs in MC clusters (typically [M/H] ~ –0.5 dex). Does this imply a correlation between the stellar rotation distribution and the stellar density of their environment? Based on the most recent *Gaia* high-precision astrometry and photometric data sets, it has been unveiled that OCs cover significantly larger spatial extents than previously assumed [118]. These clusters showcase expansive structures, many of which defy conventional explanations solely based on dynamical interactions influenced by the gravitational forces of the Milky Way (e.g., [119]). This may suggest that a significant fraction of the cluster mass already existed in extended halo-like structures with densities intermediate between the Galactic field and the cluster core region. What is the rotation distribution of stars in these extended structures? Do they also exhibit a bimodal distribution similar to that in the cluster core? Investigating these topics will deepen our understanding of the relationship between stellar rotation and their surrounding environment, potentially addressing the correlation between stellar rotation and stellar dynamical interactions.

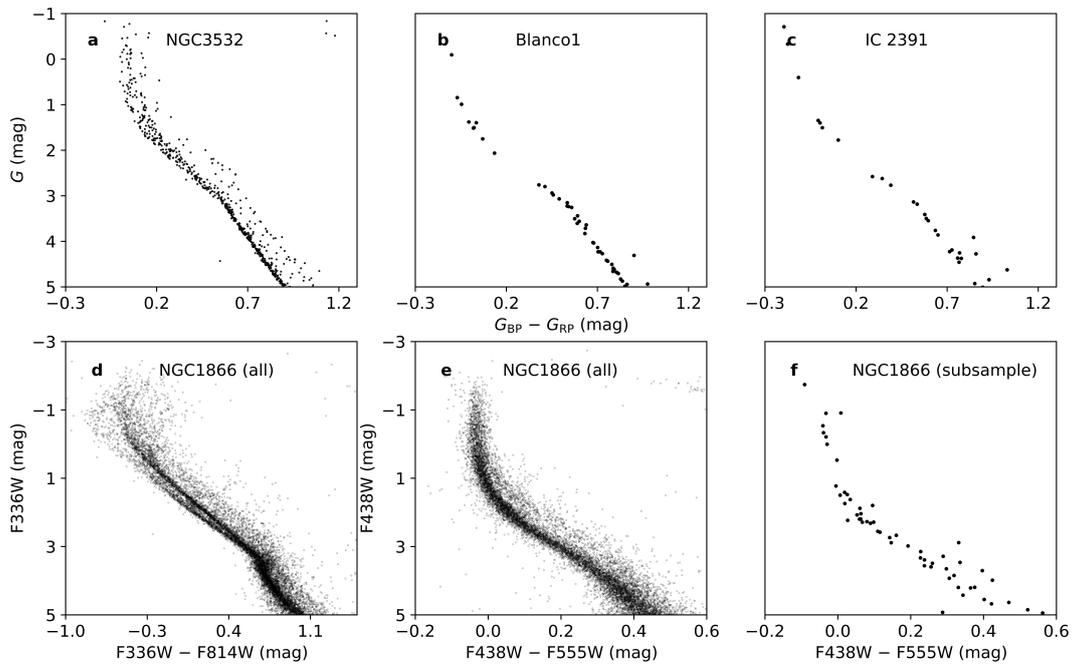

**Figure 8. Galactic OCs with (black dots) and without (blue dots) apparent eMSs.** They represent NGC 3532 (panel a, exhibiting a prominent eMS), Blanco 1 (panel b, lacking an eMS), and IC 2391 (panel c, also lacking an eMS). The bottom panels depict CMDs of NGC 1866, observed in UV–optical passbands (panel d) and optical passbands (panel e and f). The panel f showcases a CMD of a subset of NGC 1866, comprising approximately 200 stars.

Note that when we refer to the bimodal rotation distribution of stars, we are actually referring to the projected rotational velocities, $v \sin i$. Therefore, it is possible that for star clusters exhibiting a bimodal $v \sin i$ distribution, their rotation distribution is a uniform or Gaussian-like distribution, but their stellar inclinations may be bimodal. Even though the examination of NGC 2287 indicated a bimodal distribution of $v \sin i$, it was postulated that this phenomenon is primarily driven by rotation rather than inclination. This inference was drawn by comparing the observed cumulative distribution of $v \sin i$ with simulated models [16]. Direct evidence would require asteroseismic observations of these stars, which requires long-term time-domain photometric observations, however. Currently, direct asteroseismic observations have been conducted of member stars of a few OCs in the *Kepler* field, such as for stars in NGC 6791 and NGC 6819 [120]. However, these clusters are too old to exhibit eMS(TO)s, and so they cannot be used to determine the origin of the bimodal $v \sin i$ distribution. Future studies using long-term observation data from missions like the Transiting Exoplanet Survey Satellite (*TESS*) or the Zwicky Transient Facility (ZTF) can study more star clusters with eMS(TO)s and thus determine the actual rotation velocity distributions of their stars.

## 4. Summary

In this article, we have collected and reviewed studies focusing on eMSTOs and eMSs (and, sometimes, split MSs) in star clusters. We started by highlighting the traditional understanding that star clusters are composed of stars of the same age and a highly homogeneous chemical composition. However, the presence of eMSTOs and eMSs challenges this notion, apparently suggesting the coexistence of stars with different ages and compositions within a cluster.

Subsequently, this article has introduced the most significant observational characteristics of eMSs, which can be summarized as follows:

1, eMSs are observed in star clusters younger than approximately 600 Myr and are characterized by a broadening or split of the upper MS region in the CMD. eMSTOs are commonly observed in intermediate-age star clusters with ages ranging from 600 Myr to 2 Gyr. They are characterized by a broadening or extension of the MSTO region in the CMD. These features are not observed in older star clusters.

2, The presence of eMSs and eMSTOs is not related to the mass of the star cluster itself. It is primarily dependent on the masses of the individual stars within the cluster. The broadened features of eMSs and eMSTOs only manifest themselves in the range of MS masses exceeding a critical threshold.

3, The eMSTO features of star clusters can occasionally be inconsistent with other characteristics in the CMD, such as the RC and SGB. Some star clusters with eMS(TO)s exhibit a relatively narrow SGB or are accompanied by a compact RC or MSTOn. Hence, the best-fitting parameters derived from the eMS(TO)s may not coincide with those of other regions within the star cluster.

4, Young star clusters usually contain a larger fraction of Be stars compared with the average found among stars in the Galactic field. These Be stars also display redder colors in the CMD compared with normal stars.

5, For intermediate-age clusters, stars in the bright section of the eMSTO exhibit a more centrally concentrated spatial distribution than stars in the faint section. For young clusters, the spatial distributions of bMS and rMS stars differ from one cluster to another, but typically bMS stars are not more centrally concentrated than rMS stars or unresolved binaries of comparable luminosities.

6, Star clusters displaying eMSTOs or eMSs do not show significant variations in chemical abundances among their member stars, which is different from Galactic GCs containing MPs.

7, Differences in the proportion of eMS(TO)s are evident between Galactic OCs and young MC clusters, as well as in the significance of the eMSs in their CMDs.

We next introduced a range of scenarios and hypotheses that have been proposed to explain the observed eMSTOs and eMSs in star clusters. These models include a putative age spread, binary interactions, contamination by variable stars, and rapid stellar rotation. "Age spreads" refer to star clusters exhibiting prolonged star formation, including those formed through mergers of clusters or with GMCs. Binary interactions involve the production of younger stars through binary mergers or mass transfer. Contamination by variable stars refers to variable stars located in the MSTO region showing large variations in their luminosities. Rapid stellar rotation suggests that stars with different rotation velocities result in different magnitudes and colors. Each of these scenarios has its own strengths and weaknesses. To date, the most promising and successful model is the scenario that involves rapid stellar rotation.

Rapid stellar rotation can change the morphology of the MS and MSTO through three channels: gravity darkening, rotational mixing, and the formation of decretion disks. Reasonably assuming these three effects, the stellar rotation model can account for numerous observational phenomena that other models cannot explain. These include the narrow SGBs observed in star clusters with eMSTOs, the correlation between the width of the eMSTO and the age of the cluster, as well as the observed presence of UV-dim stars and MS zig-zag features, among others. Through spectroscopic observations of young and nearby star clusters exhibiting eMS(TO) features, it has finally been confirmed that stellar rotation indeed plays a crucial role in the eMS and eMSTO characteristics of star clusters.

In contrast to the clusters found in the MCs, the prevalence of eMS(TO)s in OCs within the Milky Way is relatively scarce, and the extent of these eMS(TO)s is also less significant. The underlying causes for this difference remain elusive. It is possible that manifestation of eMS(TO)s depends on the specific formation environment governing these clusters. It is also possible that this difference is simply driven by observational biases, such as the limited number of stars in Milky Way OCs and the lack of UV imaging observations. The forthcoming launch of the *CSST* will serve to bridge this knowledge gap, furnishing us with a wealth of observational data pertaining to the characteristics of eMS(TO)s in Milky Way OCs. Consequently, this will serve to constrain our understanding of the origins of eMS(TO)s.

Under the assumption that the stellar rotation model accounts for the observed eMSs and eMSTOs, the current central question focuses on the origin of the bimodal rotation distribution observed in some star clusters. Presently, the proposed explanation remains speculative, encompassing hypotheses such as binary tidal locking, protoplanetary disk interactions among proto-stars, and the role of blue straggler stars. Current observational and numerical simulations do not strongly support the hypothesis of tidal locking, while the hypotheses regarding protoplanetary disk interactions and blue straggler stars require further exploration, such as through observations of pre-MS stars within star-forming clusters and measurements of magnetic fields in bMS stars.

Finally, we highlight that based on our current knowledge about how stellar rotation could have a significant impact on the structure of star clusters, there are many valuable topics that are worth further exploration. For instance, studying the relationship between the magnetic braking effect of stars and the Galactic environment or metallicity, investigating the distribution of stellar rotation within the extended structure of OCs, understanding the formation of dusty disks around UV-dim stars, and examining whether there is a preferred orientation in the inclination distribution of stars within star clusters. The advancement of contemporary and next-generation observational instruments, such as ZTF, *TESS*, the Vera Rubin Observatory, and the *CSST*, will further drive the vibrant development of this field.

## Declaration of competing interest

The authors declare that they have no conflicts of interest in this work.

## Acknowledgments

We express our gratitude to Li Wang of Sun Yat-sen University for producing Fig. 7. We also extend our appreciation to the anonymous referee for their valuable input in enhancing this review. This research received funding from the National Natural Science Foundation of China (Grants 12233013 and 12073090) and the National Key R&D Program of China (2020YFC2201400). RdG acknowledges support from the Australian Research Council Centre of Excellence for All Sky Astrophysics in 3 Dimensions (ASTRO 3D), under project number CE170100013. APM acknowledges support from the MIUR PRIN 2022MMEB9W 'Understanding the formation of globular clusters with their multiple stellar generations' (PI: A. Marino).

**Electronic supplementary materials**

(Fundamental Research accepts electronic multimedia files (animations, movies, audio, etc.) and other supplementary files to be published online along with an article.

If supplying supplementary materials, the text must make specific mention of the material as a citation, similar to that of figures and tables. Please refer to the supplementary files as "online", e.g., "…as shown in the animation (Fig. S1 online)", "… additional data are given in Table S1 (online)".

For each supplementary material, please supply a concise caption describing the content of the file.

## References

[1] C. Dobbs. Giant molecular clouds: star factories of the galaxy, Astronomy and Geophysics. 54 (5) (2013) 24–30

[2] E. Rosolowsky, G. Engargiola, R. Plambeck, et al. Giant molecular clouds in M33. II. High-resolution observations. The Astrophysical Journal. 599 (1) (2003) 258–274

[3] C. J. Lada, E. A. Lada. Embedded clusters in molecular clouds, Annual Review of Astronomy and Astrophysics. 41 (2003) 57–115.

[4] A. G. Bruzual. Star clusters as simple stellar populations, Philosophical Transactions of the Royal Society of London, Series A. 368 (1913) (2010) 783–799

[5] H. Li, M. Vogelsberger, F. Marinacci, et al. Disruption of giant molecular clouds and formation of bound star clusters under the influence of momentum stellar feedback, Monthly Notices of the Royal Astronomical Society. 487 (1) (2019) 364–380

[6] K. Hollyhead, N. Bastian, A. Adamo, et al. Studying the YMC population of M83: how long clusters remain embedded, their interaction with the ISM and implications for GC formation theories, Monthly Notices of the Royal Astronomical Society. 449 (1) (2015) 1106–1117

[7] J. M. D. Kruijssen, A. Schruba, M. Chevance, et al. Fast and inefficient star formation due to short-lived molecular clouds and rapid feedback, Nature. 569 (7757) (2019) 519–522

[8] M. G. H. Krause, C. Charbonnel, N. Bastian, R. Diehi. Gas expulsion in massive star clusters? Constraints from observations of young and gas-free objects, Astronomy and Astrophysics. 587 (A53) (2016) 1–16

[9] E. Gavagnin, A. Bleuler, J. Rosdahl, R. Teyssier. Star cluster formation in a turbulent molecular cloud self-regulated by photoionization feedback, Monthly Notices of the Royal Astronomical Society. 472 (4) (2017) 4155–4172

[10] D. Yong, J. Meléndez, F. Grundahl, et al. High precision differential abundance measurements in globular clusters: chemical inhomogeneities in NGC 6752, Monthly Notices of the Royal Astronomical Society. 434 (4) (2013) 3542–3565

[11] A. F. Marino, A. P. Milone, N. Przybilla, et al. Helium enhanced stars and multiple populations along the horizontal branch of NGC 2808: direct spectroscopic measurements, Monthly Notices of the Royal Astronomical Society. 437 (2) (2014) 1609–1627

[12] E. Pancino, D. Romano, B. Tang, et al. The *Gaia*–ESO Survey. Mg–Al anti-correlation in iDR4 globular clusters, Astronomy and Astrophysics. 601 (A112) (2017) 1–10

[13] E. Carretta, A. Bragaglia, S. Lucatello, et al. Aluminium abundances in five discrete stellar populations of the globular cluster NGC 2808, Astronomy and Astrophysics. 615 (A17) (2018) 1–13

[14] A. P. Milone, L. R. Bedin, G. Piotto, et al. Multiple stellar populations in Magellanic Cloud clusters. I. An ordinary feature for intermediate age globulars in the LMC? Astronomy and Astrophysics. 497 (3) (2009) 755–771

[15] G. Cordoni, A. P. Milone, A. F. Marino, et al. Extended main-sequence turnoff as a common feature of Milky Way open clusters, The Astrophysical Journal. 869 (2) (2018) 139–154

[16] W. Sun, C. Li, L. Deng, et al. Tidal-locking-induced stellar rotation dichotomy in the open cluster NGC 2287? The Astrophysical Journal. 883 (2) (2019) 182–190

[17] A. P. Milone, A. F. Marino. Multiple populations in star clusters, Universe. 8 (7)

(2022) 1–41

[18] G. Bertelli, E. Nasi, L. Girardi, et al. Testing intermediate-age stellar evolution models with VLT photometry of Large Magellanic Cloud clusters. III. Padova results, The Astrophysical Journal. 125 (2) (2003) 770–784

[19] S. Rubele, L. Kerber, L. Girardi. The star-formation history of the Small Magellanic Cloud star cluster NGC 419, Monthly Notices of the Royal Astronomical Society. 403 (3) (2010) 1156–1164

[20] L. Girardi, P. Goudfrooij, J. S. Kalirai, et al. An extended main-sequence turn-off in the Small Magellanic Cloud star cluster NGC 411, Monthly Notices of the Royal Astronomical Society. 431 (4) (2013) 3501–3509

[21] A. P. Milone, L. R. Bedin, G. Piotto, et al. Multiple stellar populations in Magellanic Cloud clusters – III. The first evidence of an extended main sequence turn-off in a young cluster: NGC 1856, Monthly Notices of the Royal Astronomical Society. 450 (4) (2015) 3750–3764

[22] A. P. Milone, A. F. Marino, F. D'Antona, et al. Multiple stellar populations in Magellanic Cloud clusters – IV. The double main sequence of the young cluster NGC 1755, Monthly Notices of the Royal Astronomical Society. 458 (4) (2016) 4368–4382

[23] A. P. Milone, A. F. Marino, F. D'Antona, et al. Multiple stellar populations in Magellanic Cloud clusters – V. The split main sequence of the young cluster NGC 1866, Monthly Notices of the Royal Astronomical Society. 465 (4) (2017) 4363–4374

[24] C. Li, R. de Grijs, L. Deng, et al. Discovery of extended main-sequence turnoffs in four young massive clusters in the Magellanic Clouds, The Astrophysical Journal. 844 (2) (2017) 119–130

[25] C. Li, W. Sun, R. de Grijs, et al. Extended main-sequence turnoffs in the double cluster $h$ and $\chi$ Persei the complex role of stellar rotation, The Astrophysical Journal. 876 (1) (2019) 65–75

[26] A. D. Mackey, P. Broby Nielsen, A. M. N. Ferguson, et al. Multiple stellar populations in three rich Large Magellanic Cloud star clusters, The Astrophysical Journal Letters. 681 (1) (2008) L17–L20

[27] P. Goudfrooij, L. Girardi, A. Bellini, et al. The minimum mass of rotating main-sequence stars and its impact on the nature of extended main-sequence turnoffs in intermediate-age star clusters in the Magellanic Clouds, The Astrophysical Journal Letters. 864 (1) (2018) L3–L8

[28] Y. Yang, C. Li, Y. Huang, et al. At what mass are stars braked? The implications derived from the turnoff morphology of NGC 6819, The Astrophysical Journal. 925 (2) (2022) 159–168

[29] A. P. Milone, A. F. Marino, M. Di Criscienzo, et al. Multiple stellar populations in Magellanic Cloud clusters – VI. A survey of multiple sequences and Be stars in young clusters, Monthly Notices of the Royal Astronomical Society. 477 (2) (2018) 2640–2663

[30] A. P. Milone, G. Cordoni, A. F. Marino, et al. *Hubble Space Telescope* survey of Magellanic Cloud star clusters. Photometry and astrometry of 113 clusters and early results, Astronomy and Astrophysics. 672 (A161) (2023) 1–34

[31] C. Li, R. de Grijs, L. Deng. Not-so-simple stellar populations in the

intermediate-age Large Magellanic Cloud star clusters NGC 1831 and NGC 1868, The Astrophysical Journal. 784 (2) (2014) 157–169

[32] C. Li, R. de Grijs, L. Deng. The exclusion of a significant range of ages in a massive star cluster, Nature. 516 (7531) (2014) 367–369

[33] C. Li, R. de Grijs, N. Bastian, et al. The tight subgiant branch of the intermediate-age star cluster NGC 411 implies a single-aged stellar population, Monthly Notices of the Royal Astronomical Society. 461 (3) (2016) 3212–3221

[34] N. Bastian, F. Niederhofer. The morphology of the sub-giant branch and red clump reveal no sign of age spreads in intermediate-age clusters, Monthly Notices of the Royal Astronomical Society. 448 (2) (2015) 1863–1873

[35] X. Wu, C. Li, R. de Grijs, et al. First observational signature of rotational deceleration in a massive, intermediate-age star cluster in the Magellanic Clouds, The Astrophysical Journal Letters. 826 (1) (2016) L14–L21

[36] M. Cignoni, M. Tosi, E. Sabbi, et al. Pre-main-sequence turn-on as a chronometer for young clusters: NGC 346 as a benchmark, The Astrophysical Journal Letters. 712 (1) (2010) L63–L68

[37] G. Cordoni, A. P. Milone, A. F. Marino, et al. NGC1818 unveils the origin of the extended main-sequence turn-off in young Magellanic Cloud clusters, Nature Communications. 13 (4325) (2022) 1–9

[38] M. Griggo, M. Salaris, D. Nardiello, et al. Exploring the origin of the extended main-sequence turn-off in M37 through the white dwarf cooling sequence, Monthly Notices of the Royal Astronomical Society. 524 (1) (2023) 108–117

[39] N. Bastian, I. Cabrera-Ziri, F. Niederhofer, et al. A high fraction of Be stars in young massive clusters: evidence for a large population of near-critically rotating stars, Monthly Notices of the Royal Astronomical Society. 465 (4) (2017) 4795–4799

[40] A. F. Marino, A. P. Milone, L. Casagrande, et al. Discovery of extended main sequence turnoffs in Galactic open clusters, The Astrophysical Journal Letters. 863 (2) (2018) L33–L39

[41] E. L. Hunt, S. Reffert. Improving the open cluster census. II. An all-sky cluster catalogue with *Gaia* DR3, Astronomy and Astrophysics. 673 (A114) (2023) 114–144

[42] C. Bonatto, L. O. Kerber, E. Bica, et al. Probing disk properties with open clusters, Astronomy and Astrophysics. 446 (1) (2006) 121–135

[43] G. Cordoni, A. P. Milone, A. F. Marino, et al. Photometric binaries, mass functions, and structural parameters of 78 Galactic open clusters, Astronomy and Astrophysics. 672 (A29) (2023) 1–13

[44] S. Qin, J. Zhong, T. Tang, et al. Hunting for neighboring open clusters with *Gaia* DR3: 101 new open clusters within 500 pc, The Astrophysical Journal Supplement Series. 265 (1) (2023) 12–24

[45] P. Goudfrooij, T. H. Puzia, R. Chandar, et al. Population parameters of intermediate-age star clusters in the Large Magellanic Cloud. III. Dynamical evidence for a range of ages being responsible for extended main-sequence turnoffs, The Astrophysical Journal. 737 (1) (2011) 4–13

[46] Y. Yang, C. Li, L. Deng, et al. New insights into the formation of the blue main sequence in NGC 1850, The Astrophysical Journal. 859 (2) (2018) 98–106

[47] Y. Yang, C. Li, R. de Grijs, et al. The spatial distributions of blue main-sequence stars in Magellanic Cloud star clusters, The Astrophysical Journal (1). 912 (2021) 27–41

[48] C. Li, R. de Grijs, L. Deng, et al. The radial distributions of the two main-sequence components in the young massive star cluster NGC 1856, The Astrophysical Journal. 834 (2) (2017) 156–164

[49] S. Kamann, N. Bastian, C. Usher, et al. Exploring the role of binarity in the origin of the bimodal rotational velocity distribution in stellar clusters, Monthly Notices of the Royal Astronomical Society. 508 (2) (2021) 2302–2306

[50] A. Mucciarelli, E. Carretta, L. Origlia, et al. The chemical composition of red giant stars in four intermediate-age clusters of the Large Magellanic Cloud, The Astrophysical Journal. 136 (1) (2008) 375–388

[51] A. Mucciarelli, S. Cristallo, E. Brocato, et al. NGC 1866: a milestone for understanding the chemical evolution of stellar populations in the Large Magellanic Cloud, Monthly Notices of the Royal Astronomical Society. 413 (2) (2011) 837–851

[52] A. Mucciarelli, E. Dalessandro, F. R. Ferraro, et al. No evidence of chemical anomalies found in the bimodal turnoff cluster NGC 1806 in the Large Magellanic Cloud, The Astrophysical Journal Letters. 793 (1) (2014) L6–L10

[53] I. Cabrera-Ziri, J. S. Speagle, E. Dalessandro, et al. Searching for globular cluster chemical anomalies on the main sequence of a young massive cluster, Monthly Notices of the Royal Astronomical Society. 495 (1) (2020) 375–382

[54] W. S. Oh, T. Nordlander, G. S. Da Costa, et al. A high-resolution spectroscopic search for multiple populations in the 2 Gyr old cluster NGC 1846, Monthly Notices of the Royal Astronomical Society. 519 (1) (2023) 831–842

[55] M. Cadelano, E. Dalessandro, M. Salaris, et al. Expanding the time domain of multiple populations: Evidence of nitrogen variations in the 1.5 Gyr old star cluster NGC 1783, The Astrophysical Journal Letters. 924 (1) (2022) L2–L10

[56] E. Carretta, A. Bragaglia, R. G. Gratton, et al. Na–O anticorrelation and HB. VII. The chemical composition of first and second-generation stars in 15 globular clusters from GIRAFFE spectra, Astronomy and Astrophysics. 505 (1) (2009) 117–138

[57] A. D. Mackey, P. Broby Nielsen, A. M. N. Ferguson, et al. Multiple stellar populations in three rich Large Magellanic Cloud star clusters, The Astrophysical Journal Letters. 681 (1) (2008) L17–L20

[58] P. Goudfrooij, T. H. Puzia, V. Kozhurina-Platais, et al. Population Parameters of Intermediate-Age Star Clusters in the Large Magellanic Cloud. I. NGC 1846 and its Wide Main-Sequence Turnoff, The Astronomical Journal. 137 (6) (2009) 4988–5002

[59] S. Rubele, L. Girardi, V. Kozhurina-Platais, et al. The star formation history of the Large Magellanic Cloud star cluster NGC 1751, Monthly Notices of the Royal Astronomical Society. 414 (3) (2011) 2204–2214

[60] P. Goudfrooij, T. H. Puzia, V. Kozhurina-Platais, et al. Population parameters of intermediate-age star clusters in the Large Magellanic Cloud. II. New insights from extended main-sequence turnoffs in seven star clusters, The Astrophysical Journal. 737 (1) (2011) 3–20

[61] P. Goudfrooij, L. Girardi, V. Kozhurina-Platais, et al. Extended main sequence

turnoffs in intermediate-age star clusters: A correlation between turnoff width and early escape velocity, The Astronomical Journal. 797 (1) (2014) 35–55

[62] M. R. Krumholz. The big problems in star formation: The star formation rate, stellar clustering, and the initial mass function, Physics Reports. 539 (2014) 49–134

[63] C. Conroy, D. N. Spergel. On the formation of multiple stellar populations in globular clusters, The Astrophysical Journal. 726 (1) (2011) 36–48

[64] N. Bastian, J. Strader. Constraining globular cluster formation through studies of young massive clusters – III. A lack of gas and dust in massive stellar clusters in the LMC and SMC, Monthly Notices of the Royal Astronomical Society. 443 (4) (2014) 3594–3600

[65] N. Bastian, I. Cabrera-Ziri, B. Davies, et al. Constraining globular cluster formation through studies of young massive clusters – I. A lack of ongoing star formation within young clusters, Monthly Notices of the Royal Astronomical Society. 436 (3) (2013) 2852–2863

[66] C. Li, R. de Grijs, L. Deng. Stellar populations in star clusters, Research in Astronomy and Astrophysics. 16 (12) (2016) 179–197

[67] N. Bastian, F. Niederhofer, V. Kozhurina-Platais, et al. A young cluster with an extended main-sequence turnoff: confirmation of a prediction of the stellar rotation scenario, Monthly Notices of the Royal Astronomical Society: Letters. 460 (1) (2016) L20–L24

[68] A. D. Mackey, P. Broby Nielsen. A double main-sequence turn-off in the rich star cluster NGC 1846 in the Large Magellanic Cloud, Monthly Notices of the Royal Astronomical Society. 379 (1) (2007) 151–158

[69] K. Bekki, A. D. Mackey, On the origin of double main-sequence turn-offs in star clusters of the Magellanic Clouds, Monthly Notices of the Royal Astronomical Society. 394 (1) (2009) 124–132

[70] A. Dieball, E. K. Grebel, Studies of binary star cluster candidates in the bar of the LMC. II, Astronomy and Astrophysics. 358 (2000) 897–909

[71] W. Yang, X. Meng, S. Bi, et al, The contributions of interactive binary stars to double main-sequence turnoffs and dual red clump of intermediate-age star clusters, The Astrophysical Journal Letters. 731 (2) (2011) L37–L40

[72] P. Goudfrooij, T. H. Puzia, V. Kozhurina-Platais, et al. Population parameters of intermediate-age star clusters in the Large Magellanic Cloud. II. New insights from extended main-sequence turnoffs in seven star clusters, The Astrophysical Journal. 737 (1) (2011) 3–20

[73] S. C. Keller, A. D. Mackey, G. S. Da Costa. Extended star formation in the intermediate-age Large Magellanic Cloud star cluster NGC 2209, The Astrophysical Journal Letters. 761 (1) (2012) L5–L9

[74] R. Salinas, M. A. Pajkos, J. Strader, et al. The overlooked role of stellar variability in the extended main sequence of LMC intermediate-age clusters, The Astrophysical Journal Letters. 832 (1) (2016) L14–L18

[75] R. Salinas, M. A. Pajkos, A. K. Vivas, et al. Stellar variability at the main-sequence turnoff of the intermediate-age LMC cluster NGC 1846, The Astronomical Journal. 155 (4) (2018) 183–194

[76] C. E. Martínez-Vázquez, R. Salinas, A. K. Vivas. Short-period variability in the globular cluster NGC 419 and the SMC field, The Astronomical Journal. 161 (3) (2021) 120–138

[77] C. Johnston, C. Aerts, M. G. Pedersen, et al. Isochrone-cloud fitting of the extended main-sequence turn-off of young clusters, Astronomy and Astrophysics. 632 (A74) (2019) 1–11

[78] N. Bastian, S. E. de Mink. The effect of stellar rotation on colour–magnitude diagrams: on the apparent presence of multiple populations in intermediate age stellar clusters, Monthly Notices of the Royal Astronomical Society: Letters. 398 (1) (2009) L11–L15

[79] X. Che, J. D. Monnier, M. Zhao, et al. Colder and hotter: interferometric imaging of β Cassiopeiae and α Leonis, The Astronomical Journal. 732 (2) (2011) 68–80

[80] L. Girardi, P. Eggenberger, A. Miglio. Can rotation explain the multiple main-sequence turn-offs of Magellanic Cloud star clusters? Monthly Notices of the Royal Astronomical Society: Letters. 412 (1) (2011) L103–L107

[81] A. Maeder, G. Meynet. The evolution of rotating stars, Annual Review of Astronomy and Astrophysics. 38 (2000) 143–190.

[82] N. Bastian, E. Silva-Villa. Constraints on possible age spreads within young massive clusters in the Large Magellanic cloud, Monthly Notices of the Royal Astronomical Society. 431 (2013) 122–126

[83] W. Yang, S. Bi, X. Meng, et al. The effects of rotation on the main-sequence turnoff of intermediate-age massive star clusters, The Astronomical Journal. 776 (2) (2013) 112–125

[84] R. P. Kraft. Studies of stellar rotation. V. The dependence of rotation on age among solar-type stars, The Astrophysical Journal. 150 (1964) 551–570

[85] C. Georgy, C. Charbonnel, L, Amard, et al. Disappearance of the extended main sequence turn-off in intermediate age clusters as a consequence of magnetic braking, Astronomy and Astrophysics. 662 (A66) (2019) 1–7

[86] A. F. Marino, N. Przybilla, A. P. Milone, et al. Different stellar rotations in the two main sequences of the young globular cluster NGC 1818: the first direct spectroscopic evidence, The Astrophysical Journal. 156 (3) (2018) 116–125

[87] C. He, W. Sun, C. Li, et al. The role of binarity and stellar rotation in the split main sequence of NGC 2422, The Astrophysical Journal. 938 (1) (2022) 42–53

[88] N. Bastian, S. Kamann, I. Cabrera-Ziri, et al. Extended main sequence turnoffs in open clusters as seen by *Gaia* – I. NGC 2818 and the role of stellar rotation, Monthly Notices of the Royal Astronomical Society. 480 (3) (2018) 3739–3746

[89] S. Kamann, N. Bastian, T. O. Husser, et al. Cluster kinematics and stellar rotation in NGC 419 with MUSE and adaptive optics, Monthly Notices of the Royal Astronomical Society. 480 (2) (2018) 1689–1695

[90] S. Kamann, N. Bastian, S. Gossage, et al. How stellar rotation shapes the colour–magnitude diagram of the massive intermediate-age star cluster NGC 1846, Monthly Notices of the Royal Astronomical Society. 492 (2) (2020) 2177–2192

[91] A. Dotter. MESA isochrones and stellar tracks (MIST) 0: Methods for the construction of stellar isochrones, The Astrophysical Journal Supplement Series. 222

(1) (2016) 8–18

[92] C. T. Nguyen, G. Costa, L. Girardi, et al. PARSEC V2.0: Stellar tracks and isochrones of low- and intermediate-mass stars with rotation, Astronomy and Astrophysics. 665 (A126) (2022) 1–16

[93] C. Georgy, S. Ekström, A. Granada, et al. Populations of rotating stars. I. Models from 1.7 to $15\,M_\odot$ at $Z$ = 0.014, 0.006, and 0.002 with $\Omega/\Omega_{\rm crit}$ between 0 and 1, Astronomy and Astrophysics. 553 (A24) (2013) 1–17

[94] C. Georgy, S. Ekström, P. Eggenberger, et al. Grids of stellar models with rotation. III. Models from 0.8 to $120\,M_\odot$ at a metallicity $Z$ = 0.002, Astronomy and Astrophysics. 558 (A103) (2013) 1–17

[95] P. Goudfrooij, L. Girardi, M. Correnti. Extended main-sequence turn-offs in intermediate-age star clusters: Stellar rotation diminishes, but does not eliminate, age spreads, The Astrophysical Journal. 846 (1) (2017) 22–31

[96] S. Gossage, C. Conroy, A. Dotter, et al. Combined effects of rotation and age spreads on extended main-sequence turn-offs, The Astrophysical Journal. 887 (2) (2019) 199–214

[97] S. Mathis, V. Prat, L. Amard, et al. Anisotropic turbulent transport in stably stratified rotating stellar radiation zones, Astronomy and Astrophysics, 620 (A22) (2018) 1–16

[98] S. Martocchia, N. Bastian, S. Saracino, et al. On the origin of UV-dim stars: a population of rapidly rotating shell stars? Monthly Notices of the Royal Astronomical Society. 520 (3) (2023) 4080–4088

[99] A. P. Milone, G. Cordoni, A. F. Marino, et al. Hubble Space Telescope survey of Magellanic Cloud star clusters: UV-dim stars in young clusters, Monthly Notices of the Royal Astronomical Society. 524 (4) (2023) 6149-6158

[100] F. D'Antona, F. Dell'Agli, M. Tailo, et al. On the role of dust and mass-loss in the extended main sequence turnoff of star clusters: the case of NGC 1783, Monthly Notices of the Royal Astronomical Society. 521 (3) (2023) 4462–4472

[101] F. D'Antona, M. Di Criscienzo, T. Decressin, et al. The extended main-sequence turn-off cluster NGC 1856: rotational evolution in a coeval stellar ensemble, Monthly Notices of the Royal Astronomical Society. 453 (3) (2015) 2637–2643

[102] C. He, C. Li, W. Sun et al. The role of tidal interactions in the formation of slowly rotating early-type stars in young star clusters, Monthly Notices of the Royal Astronomical Society. 525 (4) (2023) 5880–5892

[103] L. Wang, C. Li, L. Wang, et al. On the origin of the split main sequences of the young massive cluster NGC 1856, The Astrophysical Journal. 949 (2) (2023) 53–62

[104] J. R. Hurley, C. A. Tout, O. R. Pols. Evolution of binary stars and the effect of tides on binary populations, Monthly Notices of the Royal Astronomical Society. 329 (4) (2002) 897–928

[105] N. Bastian, S. Kamann, L. Amard, et al. On the origin of the bimodal rotational velocity distribution in stellar clusters: rotation on the pre-main sequence, Monthly Notices of the Royal Astronomical Society. 495 (2) (2020) 1978–1983

[106] H. M. Lamm, R. Mundt, C. A. L. Bailer-Jones, et al. Rotational evolution of low mass stars: The case of NGC 2264, Astronomy and Astrophysics, 430 (2005) 1005–

[107] C. Wang, L. Norbert, S. Abel, et al. Stellar mergers as the origin of the blue main-sequence band in young star clusters, Nature Astronomy. 6 (2022) 480–487

[108] F. R. N. Schneider, S. T. Ohlmann, P. Podsiadlowski, et al. Stellar mergers as the origin of magnetic massive stars, Nature. 574 (7777) (2019) 211–214

[109] C. Korntreff, T. Kaczmarek, S. Pfalzner. Towards the field binary population: influence of orbital decay on close binaries, Astronomy and Astrophysics. 543 (A126) (2012) 1–8

[110] S. Martocchia, I. Cabrera-Ziri, C. Lardo, et al. Age as a major factor in the onset of multiple populations in stellar clusters, Monthly Notices of the Royal Astronomical Society. 473 (2) (2018) 2688–2700

[111] N. Bastian, C. Lardo. Multiple stellar populations in globular clusters, Annual Review of Astronomy and Astrophysics. 56 (2018) 83–136.

[112] C. Li, Y. Wang, B. Tang, et al. When does the onset of multiple stellar populations in star clusters occur? III. No evidence of significant chemical variations in main-sequence stars of NGC 419, The Astrophysical Journal. 893 (1) (2020) 17–24

[113] C. Li. Multiple stellar populations at less-evolved stages. II. No evidence of significant helium spread among NGC 1846 dwarfs, The Astrophysical Journal. 921 (2) (2021) 171–178

[114] A. Bragaglia, R. G. Gratton, E. Carretta, et al. Searching for multiple stellar populations in the massive, old open cluster Berkeley 39, Astronomy and Astrophysics. 548 (A122) (2012) 1–12

[115] A. Bragaglia, C. Sneden, E. Carretta, et al. Searching for chemical signatures of multiple stellar populations in the old, massive open cluster NGC 6791, The Astrophysical Journal. 796 (1) (2014) 68–84

[116] Y. Gong, X. Liu, Y. Cao, et al. Cosmology from the Chinese Space Station Optical Survey (CSS-OS), The Astrophysical Journal. 883 (2) (2019) 203–221

[117] C. Li, Z. Zheng, X. Li, et al. Searching for multiple populations in star clusters using the China Space Station Telescope, Research in Astronomy and Astrophysics. 22 (9) (2022) 1–16

[118] X. Pang, S. Tang, Y. Liu, et al. 3D morphology of open clusters in the solar neighborhood with *Gaia* EDR 3. II. Hierarchical star formation revealed by spatial and kinematic substructures, The Astrophysical Journal. 931 (2) (2022) 156–170

[119] J. Zhong, L. Chen, M. B. N. Kouwenhoven, et al. Substructure and halo population of Double Cluster *h* and *χ* Persei, Astronomy and Astrophysics. 624 (A34) (2019) 1–8

[120] E. Corsaro, Y. Lee, R. A. García, et al. Spin alignment of stars in old open clusters, Nature Astronomy. 1 (2017) 64–69